\documentclass[graybox]{svmult} 

\usepackage{mathptmx}       
\usepackage{helvet}         
\usepackage{courier}        

\usepackage{makeidx}         
\usepackage{graphicx}        
\usepackage{multicol}        
\usepackage[normalem]{ulem}	
\usepackage{hyperref}  
\usepackage{soul}   

\usepackage{physics} 
\usepackage{chemformula} 
\usepackage{array} 
\usepackage[numbers,sort&compress]{natbib} 

\newcommand{\ie}[0]{\textit{i.e.,}}
\newcommand{\eg}[0]{\textit{e.g.,}}
\newcommand{\EPOL}[0]{$E_{\rm POL}$}
\newcommand{\EEL}[0]{$E_{\rm EL}$}
\newcommand{\EST}[0]{$E_{\rm ST}$}

\newcommand{\VO}[0]{$V_{\rm O}$}
\newcommand{\cVO}[0]{$c_{V_{\rm O}}$}
\newcommand{\dirimg}{./}

\makeindex             

\begin{document}

\title*{Small polarons in transition metal oxides}
\author{Michele Reticcioli, Ulrike Diebold, Georg Kresse and Cesare Franchini}

\institute{Michele Reticcioli \at University of Vienna, Austria \at  Ulrike Diebold \at Institute of Applied Physics, Technische Universit\"at Wien, Vienna, Austria \at  Georg Kresse \at University of Vienna, Austria \at  Cesare Franchini \at University of Vienna, Austria, \email{cesare.franchini@univie.ac.at}}
%
%
\maketitle

\abstract{
The formation of polarons is a pervasive phenomenon in transition metal oxide compounds, with a strong impact  on the physical properties and functionalities of the hosting materials. In its original formulation the polaron problem considers a single charge carrier in a polar crystal interacting with its surrounding lattice. Depending on the spatial extension of the polaron quasiparticle, originating from the coupling between the excess charge and the phonon field, one speaks of small or large polarons. This chapter discusses the modeling of small polarons in real materials, with a particular focus on the archetypal polaron material TiO$_2$. 
After an introductory part, surveying the fundamental theoretical and experimental aspects of the physics of polarons, the chapter examines how to model small polarons using first principles schemes in order to predict, understand and interpret a variety of polaron properties in bulk phases and surfaces. Following the spirit of this handbook, different types of computational procedures and prescriptions are presented with specific instructions on the setup required to model polaron effects.  
}
\vspace{2mm}
\noindent
{\small Cite as:\\
\cite{Reticcioli2019chbook} Reticcioli M., Diebold U., Kresse G., Franchini C. (2019) {\sl Small Polarons in Transition Metal Oxides}. In: Andreoni W., Yip S. (eds) {\sl Handbook of Materials Modeling}. Springer, Cham.\\
\href{https://doi.org/10.1007/978-3-319-50257-1\_52-1}{DOI:10.1007/978-3-319-50257-1\_52-1}
}

\section{Introduction}
\label{sec:intro}

\begin{figure}[!ht]
\centerline{ \includegraphics[width=0.65\columnwidth,clip=true]{\dirimg 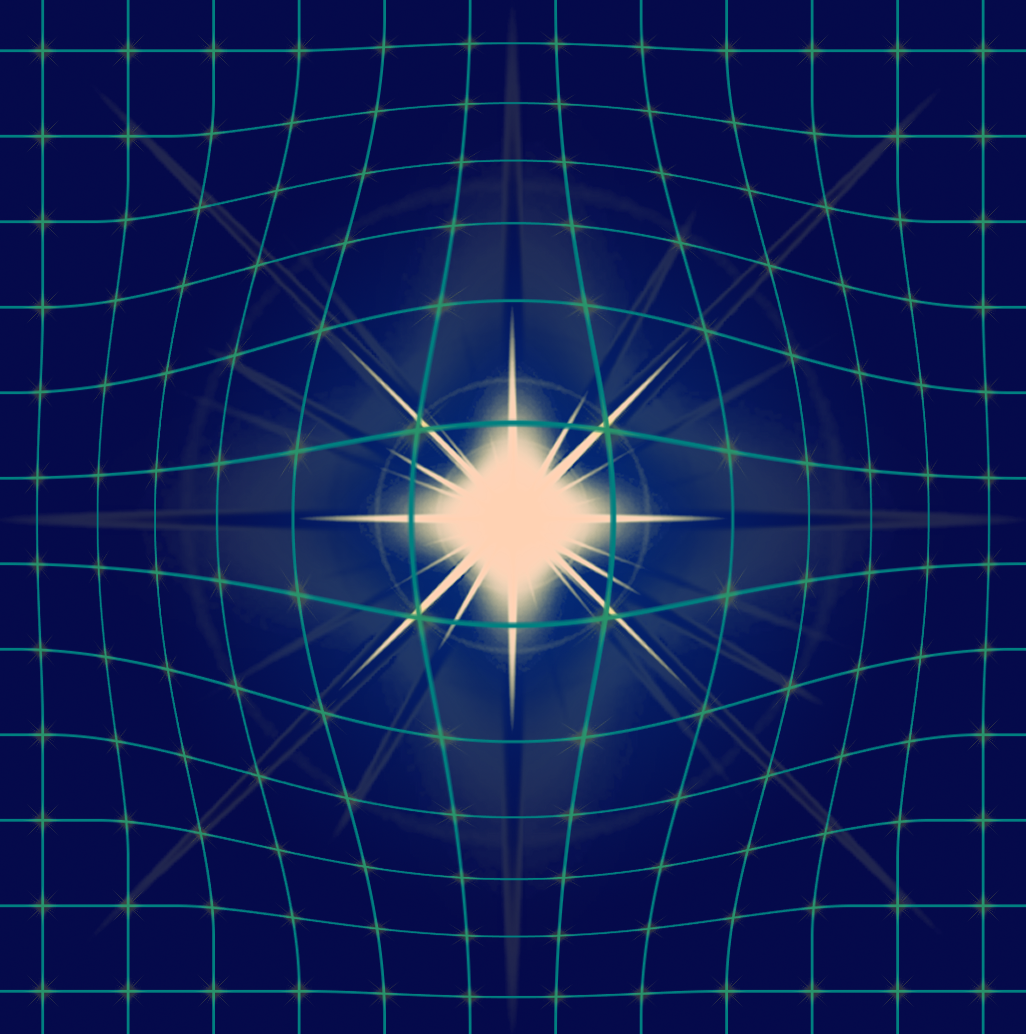} }
\caption[Schematic view of a polaron]{Pictorial view of a polaron.
An excess charge is trapped in a lattice site (bright) and distorts the surrounding lattice.
}
\label{fig:sketch}
\end{figure}

Electrons in perfect crystals are adequately described in terms of periodic wave functions.
However, rather than being constituted by a perfect lattice, real materials are often characterized by the presence of defects, such as point \hl{defects} (vacancies or interstitials), impurities, grain boundaries, and dislocations~\cite{Ashcroft,Crawford}.
Defects in the crystal break the periodicity of the electronic charge density. They can lead to the formation of localized states~\cite{Freysoldt,Pacchioni2015book}, which affect the materials properties and, thus, their performance in applications. 
Less intuitively, such localized states could exist also in the absence of any defects in the crystal, \ie\ they can form in perfect lattices.
This phenomenon is referred to as \hl{polarons}.
A polaron is a quasiparticle originating from the interactions between charge carriers (\ie\ electrons or holes) and lattice ion vibrations~\cite{Holstein1959}.
More precisely, due to Coulomb forces, the \hl{excess charge} displaces the ions in its neighborhood creating a polarization cloud that follows the charge carrier as it moves through the crystal (see Fig.~\ref{fig:sketch}).
Such quasiparticles differ from lattice defect states own peculiar properties and are described by well developed quantum field theories based on effective Hamiltonians~\cite{Frohlich1954,Holstein1959}.

Formation of polarons is particularly favorable in polar semiconductors and \hl{transition metal oxides} owing to the strength of the \hl{electron-phonon interaction} and is further promoted in the vicinity of a surface, where the crystal lattice is more flexible and the necessary lattice relaxations cost less energy~\cite{Deskins2009,Kowalski2010,Deskins2011, Setvin2014, Setvin2014b}.
Polarons play a decisive role in a wide range of phenomena, including electron transport~\cite{Crevecoeur1970,Moser2013}, optical absorption, and chemical reactivity, and have crucial implications in high-$T_{\rm c}$ superconductivity~\cite{Salje}, colossal magnetoresistance~\cite{Teresa1997,Ronnow2006}, thermoelectricity~\cite{Wang2014}, photoemission~\cite{Cortecchia2017}, and photochemistry~\cite{Linsebigler1995}.

This chapter is mostly focused on the \hl{first principles modelling} of small polarons (Ch.~\ref{ch:modelling}) and its application to \hl{TiO$_2$} (Ch.~\ref{ch:application}), an archetypal, and widely studied polaron material. This introduction is complemented with a brief summary of the fundamental theoretical aspects defining the physics of polarons and with an overview of the experimental techniques used to dected polaron features in materials.

\subsection{Theoretical background}
The emergence of polaron theory can be traced back to 1933 when Landau elaborated on the possibility for \hl{lattice distortions} to trap electrons by means of an intrinsic modification of the lattice phonon field induced by the electron itself~\cite{Landau1933others}.
The resulting electron-phonon quasiparticle was later called polaron, a coupled electron-phonon system in which the polarization generated by the lattice distortions acts back on the electron renormalizing the electronic properties, for instance the effective mass.
An analogous discussion is valid for holes rather than electrons.
A first formal description of the polaron problem was published in 1946 by Pekar~\cite{Pekar1946others}, who considered a free electron interacting with lattice deformations in the continuum approximation.
Therefore, Pekar's study is limited to the case of polarons with a size larger than the lattice constant, so that the atomic discreteness is negligible~\cite{Alexandrov2010book}.
With the ionic lattice described as a polarizable dielectric continuum, Landau and Pekar showed that the 
polaron mass $m^*$ is enhanced with respect to the ``bare`` electron mass~\cite{Landau2008}, an important result that set the basis for all subsequent theories.

During the 1950s, the second quantization formalism was used to refine the description of the polaron problem in terms of quantum effective Hamiltonians including an electronic term ($H_e$), a phonon term ($H_{ph}$) \emph{and} the electron-phonon coupling term ($H_{e-ph}$).
The $e-ph$ term is of fundamental importance for understanding the polaronic states, since it takes into account the type (e.g. short- or long-range) and intensity (weak to strong) of the mutual interactions between charge carriers and lattice vibrations.
However, a complete analytic solution of the polaron Hamiltonian cannot be achieved and approximations are needed.
Fr\"ohlich~\cite{Frohlich1950} and Holstein~\cite{Holstein1959} separately addressed different aspects of the problem and proposed mathematically more tractable formulations. Fr\"ohlich theory relies on the continuum approximation and assumes long-range forces (\hl{large polarons}), whereas Holstein theory takes into account explicitly the lattice periodicity and treats the electron-phonon coupling as a short-range interaction (\hl{small polaron}). The distinction between small and large polaron is defined according to the strength of the
electron-phonon coupling (weak/strong) and the extension of the lattice distortion around the electron (large/small)~\cite{Alexandrov2010book, Rashba2005book, Hahn2018}. 
The distinct features of small and large polarons are discussed at the end of this section.

\subsubsection{Fr\"ohlich Hamiltonian}
Fr\"ohlich theory treats the situation of long-range coupling between electrons and phonons, which leads to the formation of large polarons~\cite{Frohlich1950,Frohlich1954,Alexandrov2010book} and is formalized by the Hamiltonian:
\begin{equation}
 H = \sum_\mathbf{k} \frac{k^2}{2} c_\mathbf{k}^{\dagger}c_\mathbf{k}^{\vphantom{\dagger}} +
 \sum_\mathbf{q}   b_\mathbf{q}^{\dagger}b_\mathbf{q}^{\vphantom{\dagger}}+ \sum_{\mathbf{k,q}}\left[V_{d}^{\vphantom{\dagger}}(\mathbf{q})b_\mathbf{q}^{\dagger}c_\mathbf{k-q}^{\dagger}c_\mathbf{k}^{\vphantom{\dagger}} + V^{\dagger}_{d}(\mathbf{q})b_\mathbf{q}^{\vphantom{\dagger}}c_\mathbf{k+q}^{\dagger}c_\mathbf{q}^{\vphantom{\dagger}}\right]. 
\end{equation}
where $c_\mathbf{k}^{\vphantom{\dagger}}$ and $b_\mathbf{q}^{\vphantom{\dagger}}$ are annihilation operators for a particle with wave vector $\mathbf{k}$ and a phonon with wave vector $\mathbf{q}$, respectively, and $V_{d}^{\vphantom{\dagger}}(\mathbf{q})$ is the electron-phonon coupling function for a system in $d$ dimensions, in 3 dimensions:
\begin{equation}
 V_3^{\vphantom{\dagger}}(\mathbf{q}) = i \left(\frac{2\sqrt{2}\pi\alpha}{V}\right)^{\frac{1}{2}}\frac{1}{q}.
\end{equation}
Here $V$ is the volume of the system and $\alpha$ is the dimensionless Fr\"ohlich electron-phonon coupling constant defined in terms of 
the reduced Plank constant $\hbar$, the charge carrier charge $e$, the phonon frequency $\omega$ of the long-wavelength optical-longitudinal phonon, and the material's static and high-frequency dielectric constants $\epsilon_0$ and $\epsilon_\infty$, respectively ($\epsilon_0$ includes ionic relaxation effects, whereas the ion-clamped $\epsilon_\infty$ accounts only for electronic contributions):
\begin{equation}
\alpha=\frac{e^2}{\hbar} \sqrt{\frac{m}{2\hbar\omega}} (\frac{1}{\epsilon_\infty} - \frac{1}{\epsilon_0}).
\end{equation}
In real materials $\alpha$ ranges between $\approx$ 0.02 (InSb) to $\approx$ 3.8 (SrTiO$_3$).
In the weak-coupling regime (small $\alpha$), the Fr\"ohlich Hamiltonian can be solved using perturbation theory, and the Feynmann's path integral approach provides accurate results for all coupling strengths~\cite{Feynman1955,ROSENFELDER200163}.
In the large polaron Fr\"ohlich picture the spatial extension of the large polaron is bigger than the lattice constant and the dressed electron is 
accompanied by a phonon cloud, whose density determines the size of the polaron effectve mass $m^*$ and the polaron energy~\cite{Devreese1996, Ermin}.
Recently, the Fr\"ohlich problem has been recasted within a first-principles perspective by F. Giustino, which allows for predictive non-empirical calculations of material's specific electron-phonon properties~\cite{Giustino}.

\subsubsection{Holstein Hamiltonian}
Holstein considered the polaron determined by a short-range strong-coupling regime, the so-called small polaron, spatially confined within a radius of the order of the lattice constant, described by the general Holstein Hamiltonian~\cite{Holstein1959,Ramakumar2004}:
\begin{equation}
 H = -t\sum_{<ij>} c_{i}^{\dagger} c_{j}  + \omega\sum_i b_{i}^{\dagger} b_{i} + g\sum_i n_i (b_{i}^{\dagger} + b_{i}) 
 \label{hh}
\end{equation}
where $t$ is the hopping amplitude between neighboring sites, $c_i$($b_i$) and $c_i^{\dag}$($b_i^{\dag}$) are fermionic (bosonic) creation and annihilation operators acting on the site $i$, $\omega$ the phonon frequency, $g$ the electron-phonon coupling strength, and $n_i=c_{i}^{\dagger} c_{i}$ is the fermionic particle number operator. As opposed to the Fr\"ohlich model, the Holstein model only considers electron-phonon interactions at a single lattice site, greatly simplifying the model. 
In the simplified case of a polaron trapped in a linear molecule, hopping between different molecular sites~\cite{Holstein1959,Holstein2000}, the Hamiltonian in Eq.~\ref{hh} reduces to a two-site Hamiltonian describing the electron hopping between two sites, interacting with an ion placed in between via its vibrational mode~\cite{Alexandrov2010book}:
\begin{equation}
H = t (c_1^{\dag} c_2 + c_2^{\dag} c_1 ) - \frac {1} {2M} \frac {\partial^2} {\partial x^2} + \frac {\gamma x^2} {2} + g x (c_1^{\dag} c_1 + c_2^{\dag} c_2 )
\end{equation}
with $M$ the ion mass, $\gamma=M\omega^2$ the spring constant and $x$ the linear ion displacement from the equilibrium position.
The state $\ket{\psi}$ with an extra electron introduced into the unperturbed system $\ket{0}$ and localized at the site 1 or 2 with amplitude probability $u(x)$ and $v(x)$, respectively, is written as
\begin{equation}
\ket{\psi} = \left [ u(x) c_1^{\dag} +  v(x) c_2^{\dag}  \right ] \ket{0}~.
\end{equation}
Despite the simplicity of this model, an analytic solution for the Schr\"odinger equation $H\psi=E\psi$ can be found only in special cases (e.g. ionic vibration perpendicular to the hopping direction), and one has to rely on numerical approximations or restrict the study to the nonadiabatic ($t \ll \omega$) or adiabatic ($t \gg\omega$) limits.
The modelling of small polaron within a first principles picture is addressed in Sec.~\ref{ch:modelling}.

\vspace{5mm}

At this point, the distinction between polarons and electrons trapped at lattice defect sites stands out clearly by inspecting the Fr\"ohlich's and Holstein's analysis.
First of all, as already mentioned above, polaron formation occurs even in absence of defects, \ie\ in the perfect crystal.
Moreover, polarons are not stuck to a specific lattice site, rather they can move around in the system.
Both these characteristics, \ie\ spontaneous localization and mobility, were pointed out by Holstein, who summarized the concepts in few brilliant sentences~\cite{Holstein1959}:
\begin{quote}
``Let us imagine that an electron is momentarily fixed at some point of the crystal.
As a result of electron-lattice interaction, the surrounding lattice particles will be displaced to new equilibrium positions [\dots\negthinspace] such as to provide a potential well for the electron.
If the well is sufficiently deep, the electron will occupy a bound state, unable to move unless accompanied by the well, that is to say, by the induced lattice deformation.''
\end{quote}
At variance, the physics of an electron attached to a defect is rather different, as it is typically not mobile and its characteristics strongly depend on the type of defect~\cite{Freytag2016}.

\subsubsection{A few additional considerations on small and large polarons}

Many other theoretical studies have further developed the  ideas of Fr\"ohlich and Holstein, leading to two alternative (and to some extent more advanced) descriptions of large and small polarons~\cite{Rashba2005book}.
Table~\ref{tab:largevssmall} summarizes the main distinctions between the large and the small polaron~\cite{Emin2013book}.

\begin{table}[h]
\begin{tabular}{ c | c }
Large Polaron & Small Polaron \\
\hline \hline
Polaron radius $\gg$ lattice parameter & Polaron radius      $\simeq$ lattice parameter \\
\hline
Shallow state ($\sim 10$~meV below CBM) & In-gap state ($\sim 1$~eV below CBM) \\
\hline
Coherent motion (scattered occasionally by phonons) & 
Incoherent motion (assisted by phonons) \\
\hline
Mobility $\mu \gg 1$~cm$^2$/Vs \newline & Mobility $\mu \ll 1$~cm$^2$/Vs \\
\hline
Decreasing mobility with increasing temperature & Increasing mobility with increasing temperature \\
\hline
\end{tabular}
\caption[Large versus small polaron]{General properties of large and small polarons.}
\label{tab:largevssmall}
\end{table}

\begin{figure}[!ht]
\centering
        \includegraphics[width=0.8\columnwidth,clip=true]{\dirimg 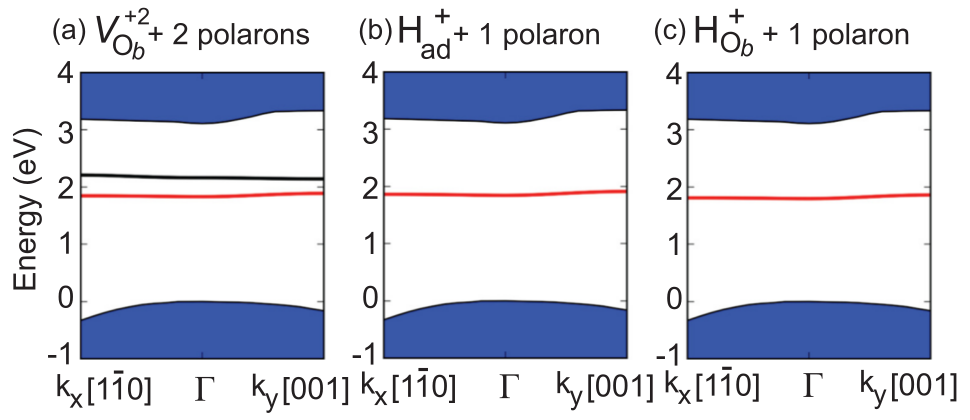}
\caption[Polaron level]{\textbf{Polaron in-gap level.} Small polaron band-gap states in the surface of rutile TiO$_2$ formed by excess electrons donated by so-called bridge-bonded oxygen vacancies ($V_{\mathrm{O}_b}$), hydrogen adsorbates (H$_{\rm ad}$), or hydrogen hydroxyls (H$_{{\rm O}_b}$).
Results based on first principles calculations within the hybrid-DFT approach.
This figure is taken from Ref.~\cite{Moses2016}.
}
\label{fig:pollevel}
\end{figure}

As already mentioned their names reflect the length scale of the spatial localization: while the small-polaron electronic charge is usually confined in the primitive cell (a few~\AA), the large polaron extends over several lattice sites (typically about 20~\AA).
Also, the two types of polaronic states exhibit distinct energy scales.
The large polaron is usually a shallow state, few tens of~meV below the conduction band minimum (CBM).
Conversely, the small polaron is firmly trapped by local distortions in a stronger potential well, which determines the formation of a deeper in-gap state well localized around 1~eV below the CBM (see Fig.~\ref{fig:pollevel}).

The localization process is also different: a charge carrier introduced into a system can quickly equilibrate with the lattice and form a large polaron, whereas the formation of a small polaron occurs only after overcoming an energy barrier~\cite{Mott1977,Lany2015}.
Finally, the two types of polarons are characterized by very different transport properties.
The large polaron, a heavy quasiparticle, is weakly scattered by phonons.
This weak scattering can compensate the large mass, resulting in a large mobility for the large polaron.
A small polaron instead hops between trapping sites with a lower mobility.
Since its motion is assisted by phonons, the mobility of small polarons increases with increasing temperature.
Both large and small polarons have been observed in several experiments, as reported in the next section, and studied by simulations and computational techniques. Historically, large polarons have been investigated mostly via effective Hamiltonians, in particular by means of variational treatments solved by Feynmann's path integral techniques, and by diagrammatic Monte Carlo~\cite{Feynman1955, Prokofev1998, Mishchenko2000, ROSENFELDER200163, Gerlach200301770,Gerlach2008, Hahn2018} approaches. First principles techniques are more suitable for the description of the small polaron, but successful attempts to address the large-polaron case have been presented in the last few years~\cite{Setvin2014, Verdi2017, Miyata2017}.

\subsubsection{How first principles modeling can be used}

The Fr\"ohlich and Holstein model Hamiltonians are simplifications of the more  general electron-phonon Hamiltonian that  is discussed
in great detail in a recent review of Giustino~\cite{Giustino}.
Giustino also explains, how the individual terms can be derived by first principles density functional theory, as briefly recapitulated below.
(i) The first term describes the one-electron band structure, and that can obviously be calculated and parameterized using density functional theory, hybrid functionals or the GW approximation.
The original Fr\"ohlich and Holstein Hamiltonian involve only a single band, with a parabolic dispersion in the Fr\"ohlich case, and a tight binding form in the Holstein case.
In the case of electron (hole) polarons, one thus needs to parameterize the conduction (valence) band.
(ii) The second term describes the vibrational frequencies of the optically active mode.
The frequencies as well as the eigenmodes can be calculated by density functional perturbation theory.
The Fr\"ohlich Hamiltonian involves only a single dispersion free phonon mode, which couples
with a wave vector dependent coupling term $q$.
The functional form used by Fr\"ohlich is exact in the long wave length limit.
The Holstein model involves a single local mode, where the local displacement couples (somehow) to the electronic bands.
(iii) The coupling between the electronic and phononic subsystem are described by the last term.
The matrix elements in Fr\"ohlich model can be exactly calculated using density functional
perturbation theory \cite{Verdi2017}.
The Holstein model is somewhat too simplistic and can not obviously be mapped onto first principles calculations.

The main issue with both approaches is two fold.
(i) In many cases, the restriction to a single band and phonon mode is a too severe a simplification,
(ii) and even for a single band and a single phonon mode the solution of the full many body
problem can be formidable.
This is the main reason why realistic modeling of polarons in real materials has hereto hardly been attempted.
Nevertheless, the single band and single phonon model might be very accurate for many simple polar semiconductors and insulators.
In this case, all one needs to determine are the effective masses, the phonon frequencies, and the dielectric constants.
Then the solution of the Fr\"ohlich model is fairly straightforward and expected to be accurate.
Furthermore, there is little reason to invoke first principles calculations for large polarons, if all materials parameters are experimentally established. 

The situation is more complicated for small polarons.
As already mentioned the Holstein model is a too simplistic approach: for a small polarons, the lattice distortions are usually large around an atom, and it is not obvious how to map the distortions to a single effective mode, furthermore the implicitly assumed harmonic approximation might not be accurate if the distortions are large.
However, small polarons often involve fairly large energy scales around 1~eV.
In this case, one might consider to treat the ionic degrees of freedom as classical particles.
In many transition metal oxides this should be a valid approximation, since the typical vibrational frequencies are small compared to the considered energy scales.
In other words, quantum fluctuations of the ions can be disregarded.
The second crucial approximation is to treat the electronic degrees of freedom using a mean field theory, such as DFT.
There is no need to discuss the accuracy of DFT, it often works exceedingly well.
Polarons though pose a particular challenge, since they involve localization of the electrons, which many density functionals fail to describe accurately because of self-interaction errors (see Sec. \ref{ch:modelling}).
Compared to the model approach the full first-principles approach is much simpler. 
One simply choses a large unit cell, adds an electron, and lets the system evolve into the groundstate.
The disadvantage compared to the model approach is about giving up on the quantum nature of the ions, and restricting the treatment of the electrons to a mean field approximation.
The advantages are, however, numerous.
The model Hamiltonians are obtained by expanding the full electron-phonon Hamiltonian around a reference groundstate to second order in the ionic displacements (recall, one can use density
functional perturbation theory to calculate the vibrational frequencies, as well as the electron-phonon coupling). 
Anharmonicities are therefore implicitly neglected in the model approach.
In this respect the simplistic, brute force approach certainly excels the model approach.

\subsection{Experimental observations of polarons}

Several years after the first theoretical predictions, experimental observations based on different techniques started to detect the presence of polarons in real materials and to study their properties. 
Nowadays, it is a common practice to complement experimental data with theoretical interpretations based on electronic structure simulations. Table~\ref{tab:exps} lists significant experimental studies revealing polaron formation in oxide materials.
This list also represents the variety of experimental techniques used to observe both small and large polarons, formed by both negative (electron) and positive (hole) excess charges in oxide materials.

\begin{table}[h!]
\begin{center}
\begin{tabular}{ | m{2.1cm} | m{1.9cm} | m{1.97cm} | m{2.5cm} | m{1.7cm} | }
Type & Material & Source & Exp. technique & Publication \\
\hline \hline
hole, small \newline polaron & UO$_{2}$ & oxidation & conductivity \newline measurement & 1963~\cite{Nagels1963} \\
\hline
hole, small \newline polaron & MnO & Li doping & conductivity \newline measurement & 1970~\cite{Crevecoeur1970} \\
\hline
electron, small polaron & CeO$_2$ & O vacancies & conductivity \newline and Seebeck & 1977~\cite{Tuller1977} \\
\hline
electron, small polaron & BaTiO$_3$ & Nb doping & EPR & 1994~\cite{Possenriede1994} \\
\hline
electron, small and large polaron & a-TiO$_2$ and \newline r-TiO$_2$ & Nb doping & conductivity and optical measurements & 2007~\cite{Zhang2007} \\
\hline
electron, small polaron & r-TiO$_2$ & O vacancies & Resonant \newline photoelectron diffraction & 2008~\cite{Kruger2008} \\
\hline
electron, small polaron & r-TiO$_2$ & O vacancies & EPR & 2013~\cite{Yang2013} \\
\hline
electron, large polaron & a-TiO$_2$ & O vacancies & ARPES & 2013~\cite{Moser2013} \\
\hline
electron, small polaron & r-TiO$_2$ & O vacancies & STM and STS & 2014~\cite{Setvin2014} \\
\hline
electron, small polaron & r-TiO$_2$ & UV irradiation \newline or H adatom  & IR spectroscopy & 2015~\cite{Sezen2015a} \\
\hline
hole, small \newline polaron & LiNbO$_3$ & Visible light & IR spectroscopy & 2016~\cite{Freytag2016} \\
\hline
electron, small polaron & r-TiO$_2$ & O vacancies & IR spectroscopy \newline on adsorbates & 2017~\cite{Cao2017a} \\
\hline
electron and \newline hole, large \newline polaron & lead halide perovskites & laser pulse & TR-OKE & 2017~\cite{Miyata2017} \\
\hline
\end{tabular}
\caption{List of experimental observations of polarons in oxide materials using different type of techniques.}
\label{tab:exps}
\end{center}
\end{table}

\subsubsection{Conductivity measurements}
The first observation of polarons is attributed~\cite{Stoneham2007} to an experimental study published in 1963~\cite{Nagels1963}.
Experiments on oxidized uranium dioxide reported an increasing hole conductivity with raising temperature, following a behavior well described by the formula:
\begin{equation}
\mu \sim \frac {1} {T} \exp(- \frac {\Delta E} {k_B T})~,
\label{eq:mu}
\end{equation}
indicating small polaron hopping from U$^{5+}$ to U$^{4+}$ sites upon thermal activation, an interpretation further substantiated by the observation that the activation energy $\Delta E$ decreases with increasing oxygen concentration.

Experimental confirmations of Eq.~\ref{eq:mu} were obtained for a wide range of materials,
with charge carriers injected by different types of defects.
Examples are given by the hole small polarons observed in Li-doped MnO~\cite{Crevecoeur1970}, and by the electron small polarons in oxygen deficient cerium dioxide~\cite{Tuller1977}.
In the latter example, two excess electrons originate from every \hl{oxygen vacancy} present in the system. Each excess electron localizes at one Ce$^{4+}$ site forming a Ce$^{3+}$ ion.
Hopping is activated by temperature, and the conductivity measurements confirmed the expected trend for polarons (Eq.~\ref{eq:mu}).
Moreover, by measuring the Seebeck coefficient for thermo-electricity, the number of carriers was found to be temperature independent.
This is in contrast with the band model for the conductivity, which predicts an increasing number of charge carriers with raising temperature.
On the contrary, according to the polaron hopping model, the charge carriers are introduced in CeO$_2$ only by defects, and the mobility increases with raising temperature due to the increased phonon populations.

\subsubsection{Electron paramagnetic resonance}
Electron paramagnetic resonance (EPR) is another experimental technique able to identify lattice ions with unpaired electrons, a distinct feature of the polaron state.
In EPR experiments, an external magnetic field splits the energy level of the unpaired electrons (Zeeman effect), thus determining energy levels available by emitting or absorbing a photon at a specific frequency: the sample is illuminated by light, typically at constant frequency, while the magnetic field varies and the
resonance peaks are measured as function of the magnetic field when the conditions for the level transitions are satisfied.
The shape, intensity and energy values of the resonance peaks determine the electronic states of the atoms in the sample where charge trapping can be detected.
Small polarons were identified by EPR in Nb doped BaTiO$_3$ samples~\cite{Possenriede1994,Lenjer2002}, and other materials including oxygen-deficient rutile TiO$_2$ samples~\cite{Yang2013}.

\subsubsection{Optical measurements}
The difference between small and large polarons manifests itself prominently in the optical properties of materials.
An example is the response of two different polymorphs of TiO$_2$, rutile and anatase, to (electron) doping with Nb, reported by Zhang \textit{et al.}~\cite{Zhang2007}.
Epitaxial thin films of the two different polymorphs were grown on appropriate substrates,  \ch{SrTiO3}(001) or \ch{LaAlO3}(001) for anatase and \ch{Al2O3}(r and c-cut) for rutile.
By substituting 4-valent Ti with 5\% of 5-valent Nb, electrons were added to the system.
The two types of films showed contrasting behavior in conductivity measurements: anatase films were metallic while rutile films semiconducting.
Together with the similar carrier density observed in both films, this is consistent with the formation of large and small electron polarons, respectively.
The optical transmittance of 80\% in the visible range for epitaxial anatase films shows that these films can be characterized as transparent conductive oxides~\cite{Klein2012}.

\subsubsection{Resonant photoelectron diffraction}

The charge distribution in reduced TiO$_2$ rutile (110) single crystals was first determined by Kr\"uger \textit{et al.}~\cite{Kruger2008,Kruger2012}.
Their experiment is based on angle-resolved x-ray photoemission (XPS).
Intensity variations in a specific photoelectron peak (here Ti-$2p$) are recorded while changing the polar and azimuthal emission angle.
Since forward-focusing dominates the scattering of electrons with kinetic energies of a few hundred eV, the configuration of near-range atomic neighbors can be accessed in a rather direct manner, and modeled in a simple cluster geometry~\cite{Chambers1992}.
In their work, the authors took advantage of the fact that the Ti-$2p$ XPS peak shows a clear shoulder that is attributed to Ti$^{3+}$, the intensity of this feature was additionally increased by tuning the photon energy to a resonance condition~\cite{Bertel1983}.
The photoelectron diffraction pattern of the Ti$^{4+}$ peak (from Ti$^{4+}$ at regular lattice sites) and the reduced Ti$^{3+}$ signature turned out to be drastically different, and the best fit was obtained by attributing the excess electrons to the subsurface Ti atoms.
This is true independently of the way the excess electrons are introduced into the lattice, either by creating O vacancies~\cite{Kruger2008} or by adding electrons via Na deposition~\cite{Kruger2012}. This points to the fact that the location of polarons in subsurface position is an intrinsic feature of rutile \ch{TiO2}.

\subsubsection{Angle resolved photoemission spectroscopy}
\ch{TiO2} has been investigated by numerous different experimental techniques in addition to the methods mentioned above. 
Angle resolved photoemission spectroscopy (ARPES)~\cite{Moser2013} was used to identify the presence of large polarons in the anatase polymorph of titanium dioxide.
The energy dispersion close to the Fermi level measured by ARPES experiment shows satellite (shallow) states below the conduction band, corresponding to large electron polarons, brought about by the oxygen vacancies present in the sample.
By tuning the amount of oxygen vacancies (via UV irradiation~\cite{Locatelli2007}), the density of charge carriers can be controlled.
At high vacancy concentrations, the satellite states disappear, due to the overlap of the polaronic wavefunctions, giving rise to a metallic behavior indicated by the crossing of the conduction band with the Fermi energy.

\subsubsection{Scanning tunneling microscopy and spectroscopy}
A direct view at the polaronic states in TiO$_2$ can be achieved by using scanning tunneling microscopy (STM) and spectroscopy (STS)~\cite{Setvin2014}.
\hl{In-gap states} stand clearly from the STM and STS studies on both anatase and rutile \ch{TiO2} polymorph, due to charge carriers (electrons) induced by oxygen vacancies or Nb doping. 
However, the two polymorphs present quite different polaronic states: while rutile \ch{TiO2} is a prototypical small-polaron material, anatase was found to host predominantly large polarons.
STM is capable to provide information on the spatial distribution of the polaronic charge at the surface, which is helpful for the identification of small and large polarons.

\subsubsection{Infrared spectroscopy}
Small polaron states have been detected in TiO$_2$ also by absorption infrared (IR) spectroscopy experiments~\cite{Sezen2015a}.
The vibrational energies of the lattice bonds are typically in the infrared regime.
Therefore, by exposure of the sample to IR light, resonance peaks appear in the measured transmitted and reflected beams.
In presence of small polarons, the localized charge together with the local lattice distortions contribute to form resonant peaks at characteristic vibrational frequencies.
Interestingly, the polaron peaks in TiO$_2$ were found to be independent on the source of charge carriers,
H adsorption and UV irradiation, a strong evidence of the polaron nature of the peaks.
The characteristic vibrational spectra of a defect state, at a donor would, in fact, be susceptible to the type of the donor.

Infrared spectroscopy can also be used to detect polarons indirectly, by inspecting the vibrational properties of bonds in the vicinity of the trapping site.
Electron small polarons were found to redshift the stretching frequency of NO molecules adsorbed on reduced TiO$_2$ surfaces with respect to the pristine samples~\cite{Cao2017a}.
Analogously, hole small polarons modify the vibrational frequency of the OH impurities at Li vacancies in \ch{LiNbO3} samples~\cite{Freytag2016}.
In the latter study, irradiation in the visible region was used to generate hole small polarons at O sites, and electron polarons at Nb sites.
The hole polarons localize in the vicinity of the Li vacancies, due to the a more favorable electrostatic potential, and therefore strongly contribute to the frequency shift of the OH impurities.
Once the source is turned off, hole and electron polarons recombine, since their mobile character, and the original vibrational frequency is restored.

\subsubsection{Time-resolved optical Kerr effect}
Recently, the time-resolved optical Kerr effect (TR-OKE) was used to investigate polaron formations in \ch{CH3NH3PbBr3} and \ch{CsPbBr3} perovskites~\cite{Miyata2017}.
In a TR-OKE experiment, a laser pulse is sent to the material and the polarization rotation is detected.
With laser pulses of energy larger than the energy-gap value, the experiment probes the TR-OKE response upon charge injection.
In these lead halide perovskites, the detected signals are compatible with a description of the altered phonon dynamics in terms of the formation of hole and electron, large polarons in the \ch{PbBr3} sublattice.
This TR-OKE study reported also interesting insights on the polaron-formation dynamics.
In fact, the two materials show different rates for the polaron formation.
This is attributed to the reorientations of the cations in \ch{CH3NH3PbBr3}, which determine a faster polaron formation than in \ch{CsPbBr3}.

\section{Modeling small polarons by first principles}
\label{ch:modelling}

The \hl{electronic structure} community has been quite active in the study of polarons in real materials.
Simulations can be helpful to understand and interpret the experimental findings or to predict the formation of polaronic states. This section provides a short overview of the first principles methods generally used to model polarons in materials and discusses different computational procedures to acquire information on the formation and dynamics of polarons and on their mutual interaction.

\subsection{Theories and methods}
\hl{Density functional theory} (DFT) has been largely applied as a starting point to study polarons in materials. However, common schemes used to describe the exchange-correlation functional in DFT simulations, such as the local density and generalized gradient approximations (LDA and GGA, respectively), fail to describe properly the charge localization at atomic sites. By using LDA or GGA, the excess charges are delocalized overall in the lattice and partial occupations of the available electronic levels is favored against integer occupations.
Therefore, the modeling of polarons requires a correction in order to overcome the drawback of standard local and semilocal exchange-correlation approximations.

\begin{figure}[!ht]
\centering
        \includegraphics[width=0.6\columnwidth,clip=true]{\dirimg 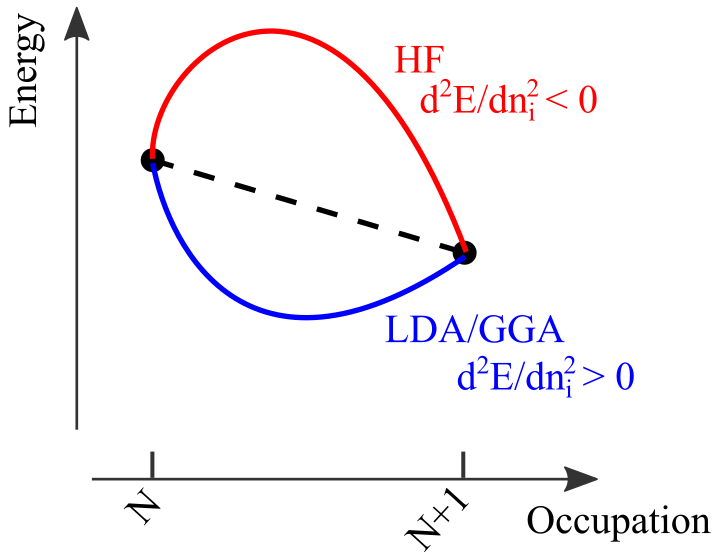}
\caption[Occupation]{\textbf{Total energy vs. occupation in DFT and HF.} 
Total energy as a function of the electronic occupation in DFT (convex function) and HF schemes (concave function).
}
\label{fig:occupation}
\end{figure}

According to the Janak's theorem~\cite{Janak1978}, variations of the total energy $E$ due to the electronic occupation ($n_i$) of the state $i$ are given in terms of the DFT Hamiltonian eigenstates $\epsilon_i$, as
${{\rm d} E} / {{\rm d} n_i} = \epsilon_i$,
independently of the exchange-correlation approximation.
The expected behavior of the exact total energy is a piecewise linear function of the electronic occupation, with discontinuities in the first derivative for integer values ($n_i=N$)~\cite{Perdew1982,Perdew1983}, that is 
${{\rm d}^2 E} / {{\rm d} n_i^2} = 0$ except at integer occupancies. 
Therefore, the energy of the state $i$ remains constant during electron addition or removal.

At variance with the expected behavior, LDA and GGA generally result  in a convex function for the total energy, \ie\ 
${{\rm d}^2 E} / {{\rm d} n_i^2} > 0$. This is schematically shown in Fig.~\ref{fig:occupation}.
The energy change of the state $i$ upon its own occupation reflects a spurious self-interaction effect, introduced by the type of exchange-correlation approximation.
As a consequence of the convexity of the calculated energy, partial occupations are preferred over integer occupations.
This leads to well known failures of DFT simulations including the underestimation of the energy band gaps, the description of strongly-correlated insulators as metals, and the difficulty to account for charge localization.

The tendency of DFT to delocalize charge can be corrected via modifications of the exchange-correlation approximations~\cite{To2005}.
An efficient remedy is to construct \hl{hybrid functionals} by mixing LDA or GGA with the Hartree Fock (HF) exchange~\cite{becke:1372}, following the general formula 
\begin{equation}
E_{XC}^{\mathrm Hybrid} = {\alpha_1}E_X^{HF} + {\alpha_2}E_X^{LDA/GGA} + \alpha_3E_C^{LDA/GGA},
\end{equation}
The mixing ratios $\alpha$'s (and in some screened hybrids the screening length) can be tuned to match physical quantities determined empirically, such as the energy band gap~\cite{He2012}.
Alternatively, the hybrid-functional parameters can be determined \textit{ab initio} by requiring the linearity of the total energy~\cite{Kronik2012,Elmaslmane2018} or following a self-consistent procedure~\cite{He2017} or fitting the parameters to the dielectric function of the material~\cite{PhysRevMaterials.2.073803,acs.jpclett.8b00919} .
This is possible since HF theory predicts an opposite result as compared to DFT:
The energy change upon electronic occupation is described by a concave function of the electronic occupation, 
${{\rm d}^2 E} / {{\rm d} n_i^2} < 0$,
thus, the charge localization is overestimated by HF.
This overestimation is due to neglecting screening effects, which are usually not negligible in solids.

As an alternative to hybrid functionals, which are usually computationally quite demanding, corrections can be applied to the standard formulation of the density functional theory in order to restore the expected behavior for the charge localization~\cite{Maxisch2006,Nolan2006a}.
To this class of methods belongs the \hl{DFT+$U$} method, where an additional term is added to the expression for the total energy.
The DFT+$U$ total energy $E_{\rm DFT+U}$ is given by
\begin{equation}
E_{\rm DFT+U} = E_{\rm DFT} + E_{\rm U}(U,J)~,
\label{eq:intro-U}
\end{equation}
where $E_{\rm DFT}$ is the energy obtained by standard DFT, while $E_{\rm U}$ is an on-site correction arising from a local Hubbard-like Coulomb repulsion ($U$) and an Hund's parameter ($J$), including double-counting corrections.
Various expressions for $E_{\rm U}$ have been proposed~\cite{Anisimov1991,Anisimov1993,Liechtenstein1995,Dudarev1998,Dudarev1998a}, such that the integer occupation of electronic states is energetically favored. 
Clearly, the results depend on the choice of $U$ and $J$,
which is not trivial, and, as a matter of fact, these quantities are typically treated as fitting parameters, by adjusting their values such that a specific quantity (e.g. the band gap) is predicted accurately.
In order to maintain the \textit{ab initio} character of DFT, procedures have been defined  to calculate the DFT+$U$ parameters from first principles such as the constrained local-density approximation~\cite{Gunnarsson1990,Cococcioni}, where the interaction parameters are obtained by considering the total-energy variation with respect to the occupation number of localized orbitals,
and the constrained random-phase approximation (cRPA), which allows for an explicit calculation of the matrix element of the screened interactions $U$ and $J$~\cite{Aryasetiawan2004}.
Alternatively, in the specific case of excess charge introduced by impurities, the $E_{\rm U}(U,J)$ term can be substituted by a on site angular dependent potential that does not affect the states of the defect-free system. The potential depends on parameters which are tuned to restore the linearity of the total energy~\cite{Lany2009b,Lany2011}.

\subsection{Polaron formation: energetics and structural distortions}
In the framework of DFT, regardless of the specific choice for the correction to the local/semi-local exchange-correlation approximation (DFT+$U$ or hybrids), the stability of polarons is analyzed in terms of a set of different energies.
Unfortunately, the notion of the \hl{polaronic energies} is not well standardized, and it varies according to the authors' preferences. 
The important ingredients to consider in order to characterize a polaron system are: 
\begin{enumerate}
 \item Type of electronic state: (a) excess charge localized in a lattice site (polaron solution with integer particle occupation on an energy level localized at one lattice site), or (b) delocalized in the entire sample (no polaron, fractional occupation on several lattice sites).
 
 \item Type of structural solution: (a) local distortions at the trapping site (polaron solution) or (b) absence of local distortions (uniform lattice, no polaron). The lattice distortions are obtained fully self-consistently through the minimization of the forces acting on the ions\footnote{To avoid any confusion, we specify that also in the non-polaron state the forces are minimized, but, since the excess charge is delocalized in the lattice, the structural changes are uniform and generally very small and become zero for very large supercells.}.
\end{enumerate}

\begin{figure}[!ht]
\centering
        \includegraphics[width=0.9\columnwidth,clip=true]{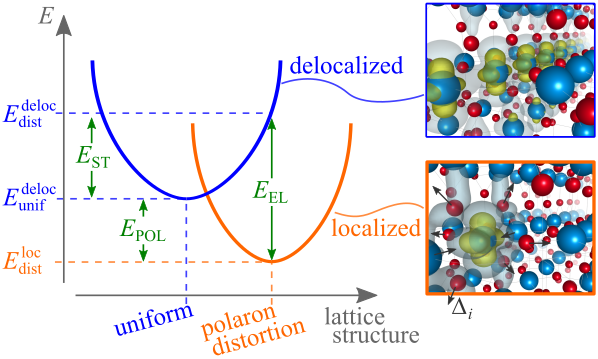}
\caption[Polaron energies]{\textbf{Polaron energies.} Sketch of the polaron formation energy (\EPOL), the structural energy cost (\EST) and the electronic energy gain (\EEL) obtained as combinations of the calculated total energies in the localized and delocalized solutions ($E^{\rm loc}_{\rm dist}$, $E^{\rm deloc}_{\rm unif}$ and $E^{\rm deloc}_{\rm dist}$).
The delocalized and localized electronic charge densities are also shown for rutile \ch{TiO2}, together with the polaronic lattice distortions $\Delta_i$.
}
\label{fig:epolsketch}
\end{figure}

Considering that the polaron represents an unpaired electron, DFT-based calculations have to be performed taking spin-polarization into account. In order to force the system to relax into a delocalized solution, it is necessary and often sufficient to use a non spin-polarized setup (this prescription is valid for both DFT+$U$ and hybrid runs).
By selectively switching on or off charge localization and/or lattice distortions, it is possible to compute the total energy of the system in different regimes (see Fig.~\ref{fig:epolsketch}):
\begin{itemize}
\item[$E^{\rm loc}_{\rm dist}$:] \hspace{5mm} Total energy of the polaron state: charge localization plus lattice distortions;

\vspace{2mm}

\item[$E^{\rm deloc}_{\rm unif}$:] \hspace{5mm} Total energy of the system with delocalized charge carriers and uniform lattice;

\vspace{2mm}

\item[$E^{\rm deloc}_{\rm dist}$:] \hspace{5mm} Total energy of the system with delocalized charge constrained into the lattice structure of the polaron state. Since
 that solution is only meta-stable, it might be difficult to realize it in practice (for instance by enforcing a non-spinpolarized groundstate).
\end{itemize}

In order to compare these energies with each other the computational setup must be kept fix (except for the spin degree of freedom) in all type of calculations, in particular the total number of electrons (including excess electrons), the unit-cell size and the value of $U$ and $J$ (or the value of the mixing parameter in hybrid-DFT).

Figure~\ref{fig:epolsketch} sketches the variations of the energy of the system in the delocalized and localized solutions as a function of lattice distortions, considering a quadratic (harmonic) energy versus structure curve.
Important insights on the formation of polarons can be obtained by combining the values of $E^{\rm loc}_{\rm dist}$, $E^{\rm deloc}_{\rm unif}$ and $E^{\rm deloc}_{\rm dist}$, which define the set of polaronic energies \EPOL (polaron formation energy), \EST (strain energy) and \EEL (electronic energy gain):
\begin{align}
E_{\rm POL}&= E^{\rm loc}_{\rm dist} - E^{\rm deloc}_{\rm unif}~,\\
E_{\rm ST}&= E^{\rm deloc}_{\rm dist} - E^{\rm deloc}_{\rm unif}~,\\
E_{\rm EL}&= E^{\rm loc}_{\rm dist} - E^{\rm deloc}_{\rm dist}~.
\end{align}
The stability of a polaron solution can be analyzed in terms of the polaron formation energy. 
A negative \EPOL\ stands for stable polarons, \ie\ the polaronic solution is energetically more convenient than a system with delocalized charge carriers.
\EST\ quantifies the structural cost needed to distort the lattice in order to accommodate the excess charge to form a polaron, whereas \EEL\ is the electronic energy gained by localizing the charge in the distorted lattice via the electron-phonon interaction.
The values of \EST\ and \EEL\ depend on the degree of charge localization and the size of the lattice distortion (see the horizontal shift of the parabolas in Fig.~\ref{fig:epolsketch}), and on the curvature of the parabola in Fig.~\ref{fig:epolsketch}.
The three polaronic energies are strongly connected.
For instance, \EPOL\  can be interpreted as the result of the competition between the structural cost \EST\ and the electronic energy \EEL~\cite{Setvin2014}:
\begin{equation}
 E_{\rm POL}= E_{\rm EL} + E_{\rm ST}~.
\end{equation}

In addition to the polaronic energies, purely structural properties can also provide important information on the nature and extension of the polaron state. To this aim one can define the average bond-length distortion $D$ around the polaron site to quantitatively measure the degree of local structural distortions~\cite{Reticcioli2018a}:
\begin{equation}
 D = \frac{1}{n} \sum_{i=1,n} |\Delta{_i}|
\label{eq:distortions}
\end{equation}
where $\Delta{_i} = \delta_{_i}^{\rm dist} - \delta_{_i}^{\rm unif}$ is the change of the bond-length $\delta$ for the ion site $i$ between the polaronic (distorted lattice) and uniform solutions; $n$ indicates the total number of ions considered in the sum (for small polarons typically first and second nearest neighbors).

\subsection{Site-controlled localization}
\label{sec:site}

Typically, \hl{charge trapping} occurs at different sites of a given material.
The various polaronic configurations are characterized by different energies.
In general, there is no guarantee that a DFT+$U$ or hybrid-functional DFT calculation leads to the global minimum of the polaronic system.
In fact, the formation of polarons could spontaneously occur at less favorable lattice sites, or occur not at all.
Therefore, it is important to control the site localization of polarons, by inspecting the formation of polarons at different sites and compare the relative formation energies. For this purpose it is essential to establish a protocol capable to selectively control charge trapping at specific sites.

Since charge localization strongly depends on the initial conditions (input) of the calculation, a selective charge trapping can be achieved by forcing initial perturbations in form of structural distortions or strong on-site Coulomb energy~\cite{Deskins2011, Shibuya2017b, Hao2015a}.
Starting from a biased setup, it is easier for the system to relax into one desired configuration at the end of the electronic \emph{and} structural self-consistent calculation (output).
Structural perturbations can be introduced manually by distorting the local structure around a given atomic site, resembling the expected polaron-induced lattice distortions. 
Alternatively, the initial perturbation can be achieved by chemical substitution, as explained below for the small electron-polaron case:
\begin{enumerate}
 \item[Step 1:] Chemical substitution at the selected site where charge trapping should occur with an atom containing one more electron than the original native atom (\eg\ Ti~$\rightarrow$~V).
Structural relaxation performed at spin-polarized DFT+$U$ level will yield local lattice distortions around the chosen site.
 \item[Step 2:] The initial element is reinserted at its original position (\eg\ V~$\rightarrow$~Ti).
It is often necessary to use a larger value of $U$ at the selected site, while the other atoms keep the original value of $U$. 
The manual initialization of the local magnetic moment has to take into account the presence of the localized electron at the selected site.
A new relaxation is performed starting from the optimized structure obtained in Step 1.
The self consistent run should be able to maintain the polaron solution obtained in Step 1.
 \item[Step 3:] A final step is necessary, performed by using the original value $U$ for all atoms. In this case it is recommended to initialize the orbitals with the one obtained in Step 2. In fact, by using a random initialization it can happen that the self-consistent loop will end up in a different polaron solution (different polaron site) or in a delocalized solution~\cite{Papageorgiou2010a,Deskins2009,Meredig2010a,Morgan2009}.
\end{enumerate}
The effective final localization of the electron at the end of each step can be verified by analyzing the local magnetic moment at the selected site.
Step 2 (\ie\ using a larger $U$) usually helps to localize the electron at the selected site.
However, it is possible, in simple problems, to skip this step and obtain charge localization at the selected site by simply performing steps 1 and 3.
This strategy can be extended to the hybrid-DFT level, by using the orbitals and the optimized structure obtained in step 3 as an input for the hybrid functional calculations.
In case more than one electron needs to be localized at selected sites, the steps 1 to 3 can be performed for every selected site separately, one after the other, or, alternatively, at the same time.
A systematic use of this strategy allows the identification of the polaronic ground state (global minimum) of the system~\cite{Reticcioli2018a}.

\subsection{Polaron dynamics}

As mentioned in the introduction, polarons are mobile quasiparticles, whose mobility can be activated or increased by temperature. Polaron hopping can be modeled in DFT, by merging first principles total energies to the Marcus~\cite{MARCUS1985265} and Emin-Holstein-Austin-Mott (EHAM) formalisms~\cite{Deskins2007a, Brunschwig,Spreafico2014e} or, more effectively, by performing \hl{first-principle molecular dynamics} (FPMD)~\cite{Kresse1993,Kowalski2010,Setvin2014} or nudged elastic band (NEB) calculations~\cite{Henkelman2000, Bondarenko2015,Janotti2013a}.

FPMD data can reveal important insights into polaron dynamics, including site population analysis (sites visited by the polarons during their hopping motion), temperature and/or concentration dependent polaron-related transitions, and polaron diffusion trajectories.
By means of a detailed \hl{statistical analysis}, one can acquire information on the interaction with specific lattice sites, with lattice or chemical defects, or on the mutual polaron-polaron interactions~\cite{Kowalski2010,Setvin2014,Hao2015a,Reticcioli2018a}. 
\begin{figure}[!ht]
\centering
        \includegraphics[width=0.8\columnwidth,clip=true]{\dirimg 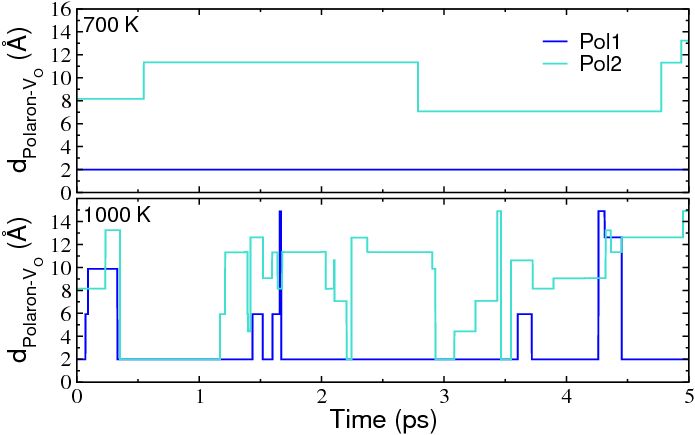}
\caption[Polaron dynamics]{\textbf{Polaron dynamics.} Example of small polaron dynamics in reduced bulk SrTiO$_3$.
Two excess electrons donated by one oxygen vacancy form two polarons whose mobility is temperature dependent: almost immobile at 700~K and  polaron-hopping at 1000~K. The position of the polarons is given in terms of their distance to the oxygen vacancy, which remains at rest at any temperature. 
Figure taken from Ref.~\cite{Hao2015a}.
}
\label{fig:poldyn}
\end{figure}
An example of polaron dynamics in oxygen deficient SrTiO$_3$ is given in Fig.~\ref{fig:poldyn}, showing the temperature-dependent interaction between two polarons and the (immobile) oxygen vacancy (\VO): at T=700~K one polaron remains anchored to \VO\ and the second one has a reduced hopping activity; at higher T (T=1000~K) both polarons become very mobile and starts to explore sites far away from \VO, but the strong Coulomb attraction between the negatively charged polaron and the positively charged \VO\ regularly brings polarons back into the vicinity of \VO, forming polaron-\VO\ complexes~\cite{Hao2015a}.

The statistical analysis is usually performed by investigating the occurrences of charge trapping at each lattice site in the simulation  cell (site-decomposed population analysis), and by computing different types of correlation functions aiming to describe the interaction of a polaron with other polarons or lattice defects:

\emph{(i) Polaron-polaron site correlation function.} The polaron-polaron site correlation function $S_{\rm pol-pol}$ can be defined as the distribution of the site distance $i$ along a particular direction in the crystal, between two polarons at a given time-step $t$, averaged over all the FPMD time steps $\tau$:
\begin{equation}
S_{\rm pol-pol}(i)= \frac 1 {N} \frac 1 {\tau} \sum_{t=0}^{\tau} \sum_{j} \rho_j(t) \rho_{j+i}(t)~,
\end{equation}
where $N$ is the number of atomic sites, and $\rho_j(t)$ indicates the polaronic site density at time $t$, and it is equal to 1 for the $j$-th site hosting a polaron, and 0 otherwise.

\emph{(ii) Radial correlation function} Analogously, the polaron-polaron $R_{\rm pol-pol}$ and defect-polaron $R_{\rm def-pol}$ radial correlation functions can be defined as a function of the distance $r$:
\begin{equation}
\begin{split}
      R_{\rm pol-pol}(r)&=  \frac 1 {\tau} \sum_{t=0}^{\tau} \sum_{(q,p)} \delta(|\mathbf{r}_q-\mathbf{r}_p|,r,t)~,\\
R_{\rm def-pol}(r)&=  \frac 1 {\tau} \sum_{t=0}^{\tau} \sum_{({\rm def},p)} \delta(|\mathbf{r}_{\rm{def}}-\mathbf{r}_p|,r,t)~,
\end{split}
\end{equation}
where the variables $\delta(|\mathbf{r}_{\rm{def}}-\mathbf{r}_q|,r,t)$ and $\delta(|\mathbf{r}_q-\mathbf{r}_p|,r,t)$ assume the value 1 if, at time $t$, the polaron $p$ is at distance $r$ from the defect at position $\mathbf{r}_{\rm{def}}$ or from the polaron $q$ at position $\mathbf{r}_q$, respectively, and are 0 otherwise.

These statistical quantities ($S_{\rm pol-pol}$ and $R_{\rm pol-pol}$) furnish information on the spatial distribution of polarons in the lattice, by indicating the statistically more favorable polaron configurations.

The FPMD calculation is also useful to investigate the stability of the polarons under different conditions, as an alternative to the site-controlled localization strategy discussed above.
In fact, a FPMD run leads to a set of high-temperature structures, hosting polarons in various configurations, which might be difficult to access by simple structural relaxations.
A useful strategy is to use the high-T structures obtained by FPMD as starting structures for relaxations, and check their stability by inspecting the corresponding polaron formation energies \EPOL\, and ultimately determine the global minimum (polaronic ground state)~\cite{Setvin2014,Reticcioli2017c}. This strategy is obviously similar
to simulated annealing.

\begin{figure}[!ht]
\centering
        \includegraphics[width=0.8\columnwidth,clip=true]{\dirimg 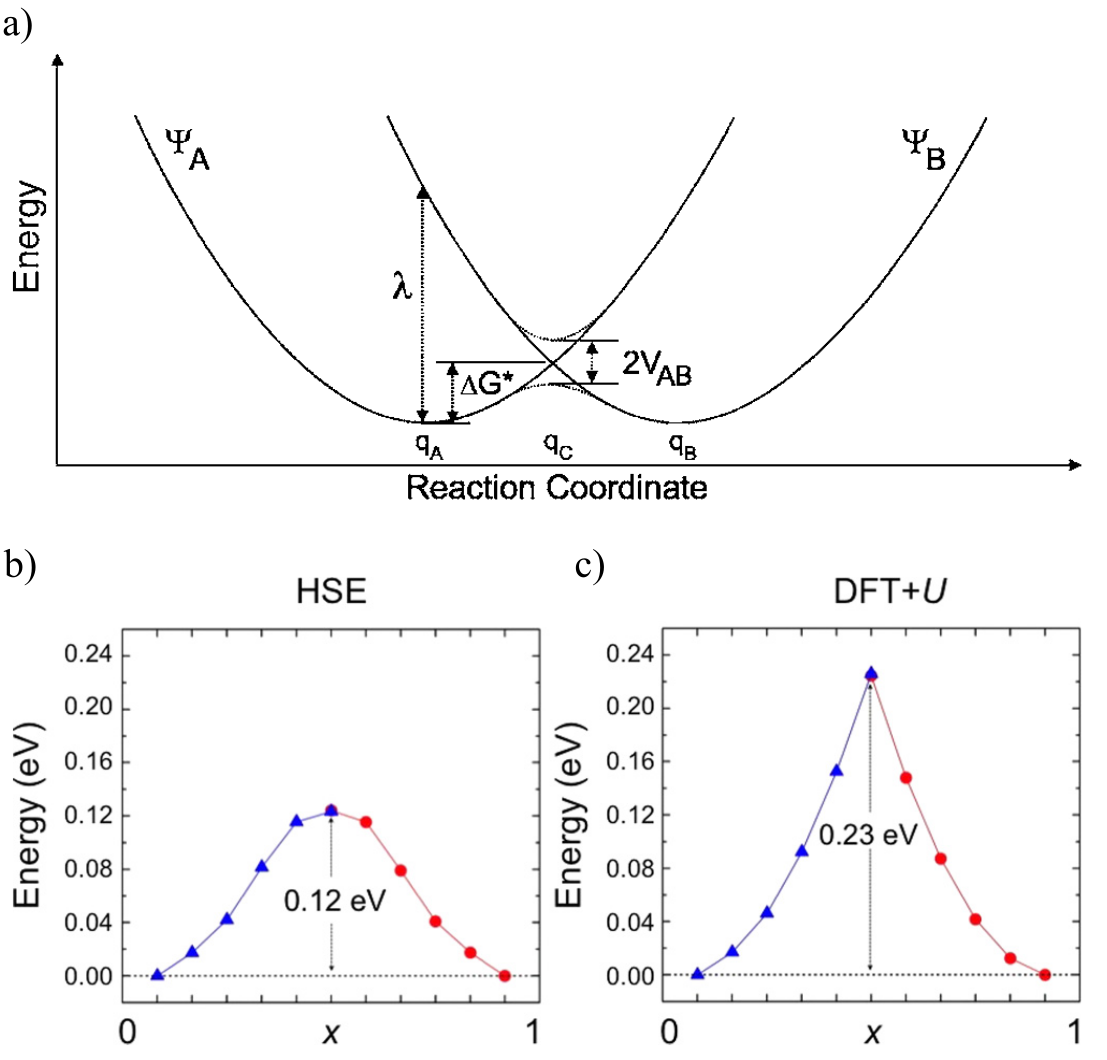} 
\caption[Polaron hopping]{\textbf{Polaron transfer.}
(a) General scheme of the Marcus-Emin-Holstein-Austin-Mott theory showing the adiabatic and the non-adiabatic mechanism, taken from Ref.~\cite{Deskins2007a}.
First principles results of polaron transfer in  \ch{CeO2} based on (b) hybrid-DFT (HSE parametrization) and (c) DFT+$U$. Depending on the computational scheme, the polaron transfer is described as an adiabatic (HSE) or non-adiabatic (DFT+$U$) process, resulting in different activation barriers.
This Figure is adapted from Ref.~\cite{Sun2017}.
}
\label{fig:hopping}
\end{figure}

As mentioned above, in addition to the FPMD approach, polaron dynamics can be investigated by a sequence of static calculations along pre-defined trajectories, using a linear interpolation scheme (LIS)~\cite{Deskins2007a,Janotti2013a} or the NEB approach. 
In the LIS scheme, the energy barrier for the transition of a polaron between different polaronic configurations can be estimated by static DFT calculations on intermediate distorted structures~\cite{Deskins2007a}.
Such intermediate structures $q(x)$ are obtained by a linear interpolation of the ionic positions assumed at the initial ($A$) and final ($B$) polaronic states, \ie\ $q(x) = (1-x)q_A + xq_B$, which define the polaron pathway from $A$ to $B$~\cite{Deskins2007a}, see Fig.~\ref{fig:hopping}.
At each considered configuration, an electronic self-consistent calculation is performed in order to compute the total energy at each step, and to construct an energy diagram of the polaron transition as a function of the reaction coordinate $x$, as those shown in Fig.~\ref{fig:hopping}~\cite{Sun2017}. The form of the energy profile determines whether the polaron transfer is an adiabatic or diabatic (\ie\ non-adiabatic) process [Fig.~\ref{fig:hopping}(a)]~\cite{Deskins2007a}. If the coupling between the initial and final state $V_{AB}$ is strong [large $V_{AB}$, see Fig.~\ref{fig:hopping}(a)] the activation barrier for polaron transfer decreases and the energy curve exhibits a smooth transition: this corresponds to an adiabatic mechanism which occurs via thermal hopping. The alternative case is when the interaction between the initial  and final  states is low and the transfer follows a non-adiabatic process. In this case the energy curve exhibits a cusp at the intermediate state and the electron transfer occurs via quantum tunneling.

It is important to note that, unfortunately, different functionals can lead to a different description of the polaron transfer process. For instance  DFT+$U$ and hybrid functional approaches can lead to qualitative different results, as shown in Fig.~\ref{fig:hopping}~\cite{Sun2017}.
In DFT+$U$, the partial localization is unfavored, the charge carrier is strongly localized in one site only, the interaction between the initial and final configuration is weak and the hopping occurs typically diabatically. In this case the energy as a function of the lattice distortions resembles the intersection of two parabola, as shown in Fig.~\ref{fig:hopping}(c).
The cusp is most likely an artifact of the one-center terms in the DFT+U approach, which does not allow for localization of the electron between two lattice sites.
Conversely, hybrid functionals predict usually an adiabatic hopping, with a more realistic gradual transfer of the electronic charge from one trapping site to the other one, as shown in Fig.~\ref{fig:hopping}(b).
As a consequence, activation barriers for polaronic hopping are usually smaller in hybrid-DFT compared to DFT+$U$. 


\section{Small polarons in \ch{TiO2}}
\label{ch:application}

Charge trapping and the formation of polarons is a pervasive phenomenon in transition metal oxide compounds, in particular, at the surface.
As already outlined, polaron formation influences the fundamental physical and chemical properties of materials, manifested by a local alteration of the bond lengths, a change of the formal valence at the specific polaronic site, and the emergence of a characteristic peak localized in the gap region~\cite{Nagels1963,Verdi2017,Sezen2015a,Freytag2016}. These changes affect virtually all functionalities of the material in practical applications.
This section is devoted to the presentation of selected results on polaron effects in TiO$_2$, one of the most studied transition metal oxides and a prototypical polaronic material.
The results are obtained using the array of first principles schemes described in the previous section. Considering that small polarons show similar properties among transition metal oxides, the methodologies adopted for TiO$_2$ should be representative for this class of compounds.

\ch{TiO2} has been studied by several experimental techniques, making the basic properties of this material well characterized~\cite{Diebold2010a}.
\ch{TiO2} is used in many applications, primarily as white pigment in paints and cosmetic products,
as physical blocker for ultraviolet UVA in almost every sunscreen, it is added to cement and tiles (to give the material sterilizing, deodorizing and anti-fouling properties), it is utilized in self-cleaning glasses as optical and corrosion-protector (this type of glass is coated in a thin layer of transparent anatase), it is a component of Gr\"atzel cells (in nanostructured form), and it can be used as varistor in electric devices.

The first principles modeling of polarons is performed by introducing excess charge carriers in a variety of ways:

\begin{enumerate}
 \item Introducing defects, such as oxygen vacancies, interstitial Ti atoms or Nb impurities.
 \item By suitably changing (by hand) the number of electrons of the system. In this way one  mimicks excess charges injected by UV-irradiation in perfect (\ie\ defect-free) samples~\cite{Sezen2015a} or cases where the donor and excess charge are spatially separated.
 \item In surface environments, by means of adsorbates (e.g. hydrogen).
\end{enumerate}

As already mentioned, the stability of the polaron state depends on many factors, in particular, the location of the trapping site (a particularly delicate issue at surfaces where the broken symmetry results in various inequivalent Ti sites), the interaction with defects and the concentration of charge carriers (polaron-polaron interaction).
The system is generally characterized by an energy profile with several local polaronic minima (even the delocalized solution might trap in a local minimum), and there is no guarantee that single-shot runs will lead to the identification of the  most favorable configuration; this issue is discussed in Sec.~\ref{sec:site}.

\subsection{Rutile and anatase}

\ch{TiO2} exists in various crystal structures (polymorphs).
Rutile and anatase, the most common and stable \ch{TiO2} polymorphs, bear the formation of polarons~\cite{Yin2018}, with some important distinctions.
The two polymorphs show different properties upon injection of an excess charge into the system, \eg\ anatase becomes metallic, while rutile remains semiconducting upon Nb doping~\cite{Zhang2007} (as mentioned in Ch.~\ref{sec:intro}), consistent with the formation of large and small polarons, respectively.
Large polarons have been reported in anatase also by ARPES~\cite{Moser2013} and STM~\cite{Setvin2014} measurements, and interpreted by first principles calculations~\cite{Setvin2014,Verdi2017}.
Conversely, small polarons have been found to localize easily in pristine or defective rutile, while their formation in anatase is more seldom.
Small polarons in anatase have been observed only in samples containing chemical or structural defects (e.g. oxygen vacancies and step edges)~\cite{Setvin2014,Setvin2014b}.

\begin{figure}[!h]
\centering
        \includegraphics[width=0.8\columnwidth,clip=true]{\dirimg 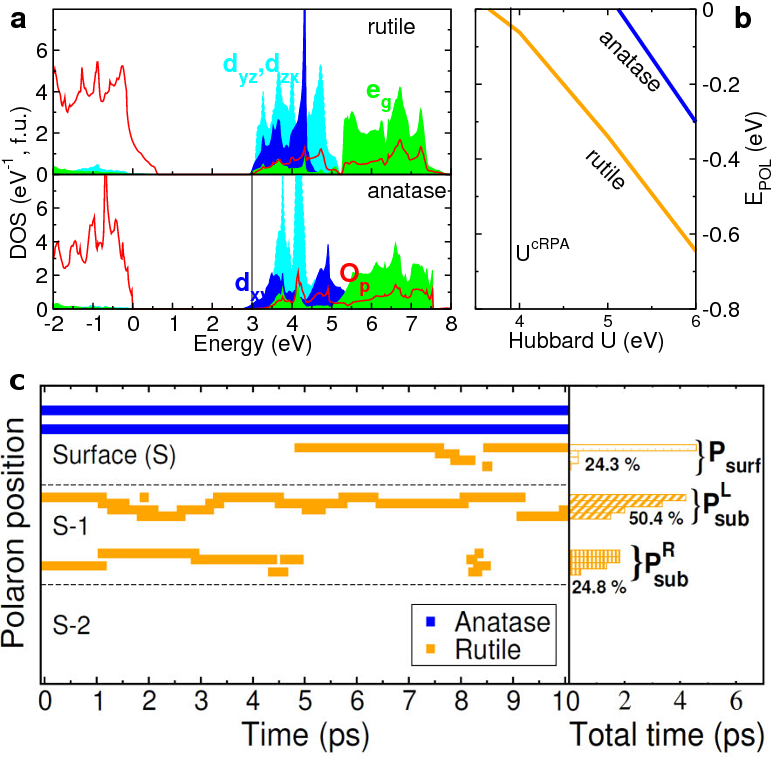}
\caption[Small polaron formation in \ch{TiO2}]{\textbf{Small polaron formation in \ch{TiO2}.} 
Basic polaron properties of rutile and anatase TiO$_2$: (a) Density of states and (b) polaron formation energies in the bulk phases and (c) FPMD trajectories for oxygen deficient surfaces.
Adapted from Ref.~\cite{Setvin2014}.
}
\label{fig:tio2u}
\end{figure}

The theoretical analysis based on first principles calculations has shed some light on this distinct behavior of rutile and anatase~\cite{Setvin2014, Spreafico2014e}. In particular it was found that the trapping energy in rutile is significantly larger than in anatase, where a larger $U$ (probably unrealistically large) is required to form a small polaron [see Fig.~\ref{fig:tio2u}(b)]. By considering a realistic $U$ of $\approx$~4~eV, obtained fully \textit{ab initio} by cRPA, small polarons are formed in rutile but not in anatase~\cite{Setvin2014}.
This behavior is explained by inspecting the density of states, shown Fig.~\ref{fig:tio2u}(a).
The formation of a polaron involves the perturbation of the conduction band minimum (CBM),
which has a different character in the two TiO$_2$ polymorphs: the CB minimum in anatase lies at a lower energy than in rutile, and it exhibits a wider bandwidth, suggesting that a larger $U$ is required to change
this favorable configuration~\cite{Setvin2014}.

On the other hand, at reducing conditions, \ie\ under conditions where O vacancies (\VO) are present, small polarons can be formed on the surface of both polymorphs, but their degree of mobility is rather different, as demonstrated by the FPMD analysis shown in Fig.~\ref{fig:tio2u}(c). In rutile, polarons (orange lines) are very mobile and hop preferentially among subsurface and surface sites, while in anatase the polaron is immobile and it stays attached to the \VO, forming a polaron-\VO\ defect-like complex~\cite{Setvin2014} (the following sections are devoted to a more detailed discussion of the small polaron properties of rutile TiO$_2$).

The representative examples discussed in this chapter are mostly focused on the rutile phase, in particular on the the TiO$_2$(110) surface.

\subsection{Small polarons on the surface of rutile TiO$_2$}

Rutile \ch{TiO2} crystallizes in a tetragonal symmetry with each Ti atom surrounded by six O atoms in a distorted octahedral configuration~\cite{Grant1959}.
Neighboring octahedra share one corner and are rotated by 90$^\circ$.
The titanium atoms are in a nominal $4+$ oxidation state (Ti$^{4+}$), while the oxygens are in a $2-$ state (O$^{2-}$).
Rutile can be easily reduced by intrinsic defects, such as interstitial Ti atoms and double-charged oxygen vacancies, and by doping and molecular adsorption.
Regardless of the source used to obtain charge carriers in \ch{TiO2}, excess electrons show similar behaviors~\cite{Janotti2013a, Deak2015b, Moses2016}: they are trapped at a titanium site, which becomes a Ti$^{3+}$ site, a hallmark revealed and confirmed by EPR experiments~\cite{Yang2009a,Yang2013}.

DFT-based simulations are capable to describe the formation of small polarons in rutile, but the stability of the polaronic states, defined in terms of the polaron formation energy, \EPOL, does depend on the computational approach and on the input parameters~\cite{Yin2018}, such as the on-site Coulomb interaction in DFT+$U$~\cite{Calzado2008,Setvin2014} and the amount of mixing in hybrid functional DFT~\cite{Spreafico2014e}.
The calculated \EPOL\ values are spread in a broad range between 0.1 and 0.8~eV~\cite{Yin2018}.
Despite these quantitative differences, the predicted properties are qualitatively in good agreement among the various methods and also with the experiments.
It has been argued that beyond DFT methods, such as RPA, could result in a more reliable quantitative determination of \EPOL\ ~\cite{Spreafico2014e}, but the high computational cost of the RPA method and the difficulty to implement forces at the RPA level~\cite{Ramberger} (in existing RPA calculations the lattice has been always relaxed at DFT+$U$ or hybrid-DFT level) have limited its applicability.
In fact, the large size of the cell, needed to accommodate the polaron together with the induced lattice distortions, prohibits simulations with more advanced quantum chemistry methods, and DFT+$U$ and to a lesser extent hybrid-DFT remain the most popular choice.

\begin{figure}[!ht]
\centering
        \includegraphics[width=0.5\columnwidth,clip=true]{\dirimg 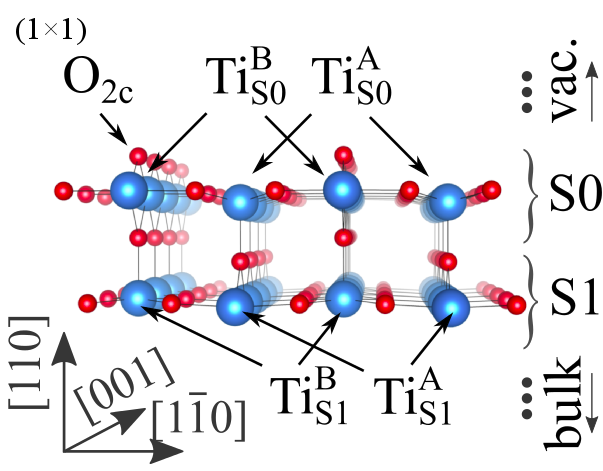}
\caption[The rutile \ch{TiO2}(110)]{\textbf{The rutile \ch{TiO2}(110).} Side view of the (1$\times$1) phase of rutile \ch{TiO2}(110).
The atoms marked by $A$ are the Ti$_{5c}$ atoms on the $S0$ layer and the atoms below; analogously, the atoms marked by $B$ are the Ti$_{56}$ atoms on the $S0$ layer and the atoms below.
This Figure is adapted from Ref.~\cite{Reticcioli2017c}.
}
\label{fig:struct}
\end{figure}

The rutile surface deserves particular attention in the study of polarons, since charge trapping is more favored in the proximity of the surface than in the bulk.
The most stable surface of rutile, the \ch{TiO2}(110) surface, consists of a bulk-terminated (1$\times$1) surface (see Fig.~\ref{fig:struct}), with large relaxations of the atoms at the surface layers, predominantly along the [110] direction.
The surface slab is formed by a sequence of tri-layer building blocks, constituted by a central layer comprising an equal number of Ti$^{4+}$ and O$^{2-}$ atoms sandwiched between two layers of oxygen atoms.
The topmost (110) layer (the surface layer) is very corrugated since it terminates with undercoordinated (twofold) oxygen atoms (O$_{2c}$) arranged in rows along [001].
The layer below the O$_{2c}$ rows is constituted by tri-fold coordinated oxygen [001] rows alternated to six-fold coordinated (Ti$_{6c}$) and five-fold coordinated (Ti$_{5c}$) titanium [001] rows, see Fig.~\ref{fig:struct}.
Obviously, the surface breaks the symmetry of the Ti and O atoms and exhibits a higher degree of flexibility and different type of structural distortions compared to the bulk-like layers.

The O$_{2c}$ atoms are relatively easy to remove by thermal annealing, irradiation, electron bombardment, or sputtering, because of their coordinative undersaturation, and this is the main source of excess charge for rutile surfaces.
In fact, the propensity to O$_{2c}$ removal reduces the system and each \VO\ results in the injection of two excess electrons, which are eligible to form small polarons.
The ability to control the amount of defects and the concentration of polarons is extremely important, as this can be used to tune the chemical and physical properties of the surface. 
For instance, under extreme reducing conditions, \ch{TiO2}(110) undergoes a structural reconstruction, from the (1$\times$1) to a (1$\times$2) phase, doubling the periodicity along [1$\bar{1}$0]~\cite{Onishi1994,Onishi1995,Onishi1996,Reticcioli2017c}.
Different models have been proposed to describe the atomic structure of this reconstruction. The debate is still open, but the \ch{Ti2O3} model proposed by Onishi appears to be the most probable one~\cite{Wang2014a, Reticcioli2017c}.

\subsubsection{Polaron configurations and properties}

\emph{Preferable trapping site}. 
The first question to address while modeling the polaron formation is the determination of the optimal polaron sites.
According to FPMD calculations, Ti sites in the subsurface layer ($S1$) are the most favorable trapping centers for excess electrons.
Specifically, the site dependent analysis of \EPOL\ shows that the most favorable sites are the Ti atoms (Ti$^A_{S1}$) at the $S1$ layer below the Ti$_{5c}$ atoms, as indicated by the histogram analysis shown in Fig.~\ref{fig:tio2u}(c).
Formation of polarons at the $S1$ (Ti$^B_{S1}$) sites below the Ti$_{6c}$ atoms is less probable, due to an unfavorable local structural relaxation.
Polaron formation at deeper layers is hindered by the stiffness of the lattice, which rises the energy cost to locally distort the lattice.
The FPMD analysis in Fig.~\ref{fig:tio2u}(c) also shows that polaron formation at the surface layer ($S0$) is not particularly favorable, mostly due to the reduced electron screening at the surface and large local distortions.

\begin{figure}[!ht]
\centering
        \includegraphics[width=0.8\columnwidth,clip=true]{\dirimg 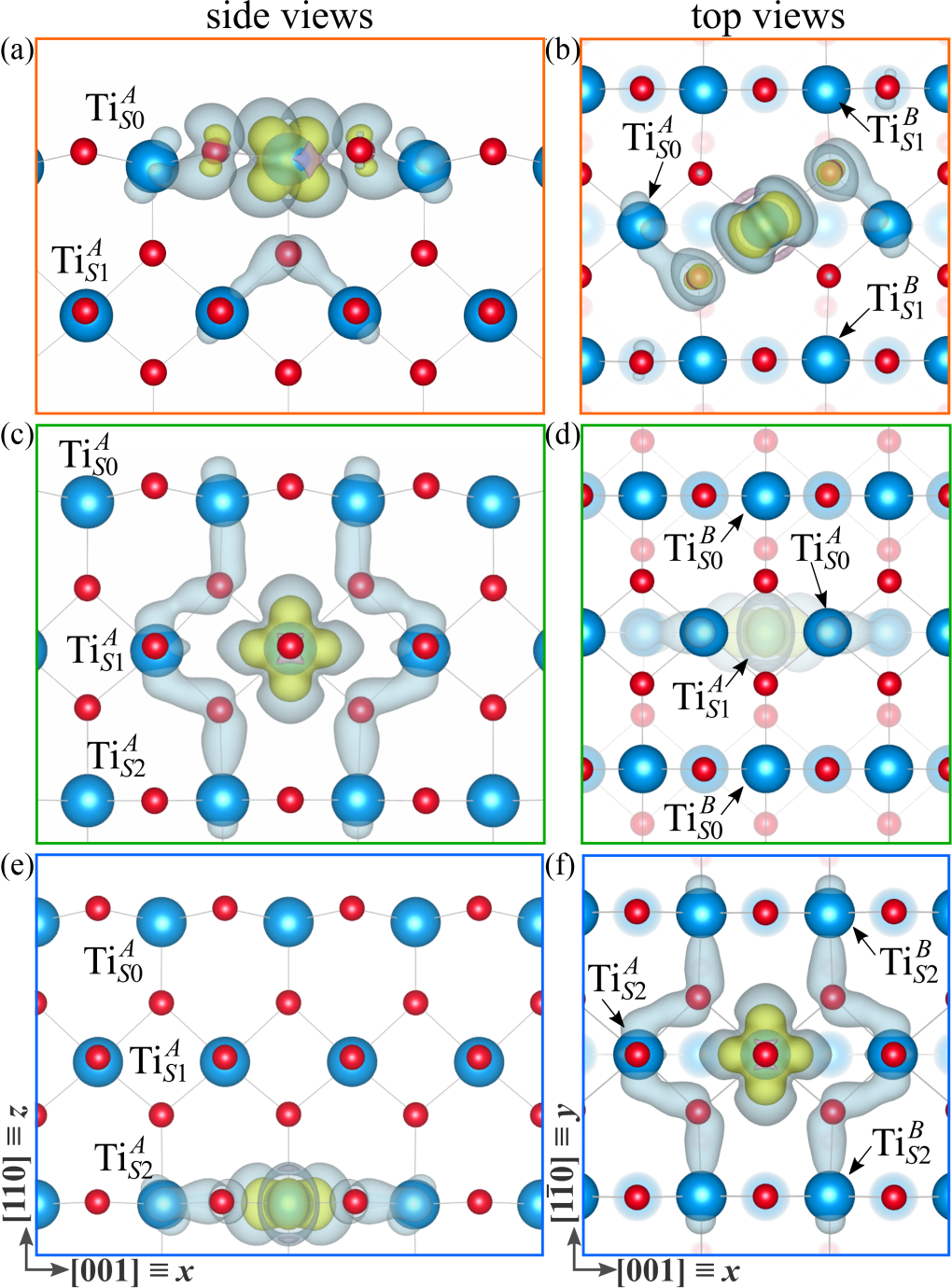}
\caption[Small polarons for the  rutile \ch{TiO2}(110) surface.]{\textbf{Small polarons for the  rutile \ch{TiO2}(110) surface.} Side (a,c,e) and top (b,d,f) views of the electronic charge density of polarons on the surface $S0$ (a,b), sub-surface $S1$ (c,d) and sub-sub-surface $S2$ (e,f) layers.
The atoms marked by $A$ are the Ti$_{5c}$ atoms on the $S0$ layer and the Ti atoms below; analogously, the atoms marked by $B$ are the Ti$_{6c}$ atoms on the $S0$ layer and the Ti atoms below.
This Figure is adapted from Ref.~\cite{Reticcioli2018a}.
}
\label{fig:clouds}
\end{figure}

\emph{Polaron orbital symmetry}. 
A second important aspect is the orbital symmetry of the polaron state.
Depending on the hosting site, the excess electron occupies different $d$ orbitals. 
Which $d$ orbital becomes populated  is predominantly determined by the local chemical environment, the crystal field splitting, and the flexibility of the neighboring atoms. 
Polarons in the Ti$_{5c}$ sites on the surface layer ($Ti^A_{S0}$) are characterized by a $d_{xz}-d_{yz}$ orbital symmetry as evidenced by the polaron charge density plot displayed in Fig.~\ref{fig:clouds}(a,b).
On the other hand, below the surface, in the subsurface layer ($S1$) and in deeper layers, small polarons assume $d_{z^2}-d_{x^2-y^2}$ and $d_{x^2-y^2}$ orbital symmetries, alternately along [110] and [1$\bar{1}$0] [see Fig.~\ref{fig:clouds}(c-f)].
This stacking is the result of the orientation of the local environment (\ie\ the orientation of the coordination octahedron around the trapping Ti site).
The lattice distortions are larger for the Ti$_{5c}$ polarons (the Ti site hosting the polaron relaxes outwards along [110], and the surrounding O atoms and the nearest-neighbor Ti atoms are pushed away by the excess electron), while the relaxations in deeper layers are smaller (the Ti-O distance in the TiO$_6$ octahedron increases, and the nearest Ti$^{4+}$ sites relax towards the polaronic site).

\begin{figure}[!ht]
\centering
        \includegraphics[width=1.0\columnwidth,clip=true]{\dirimg 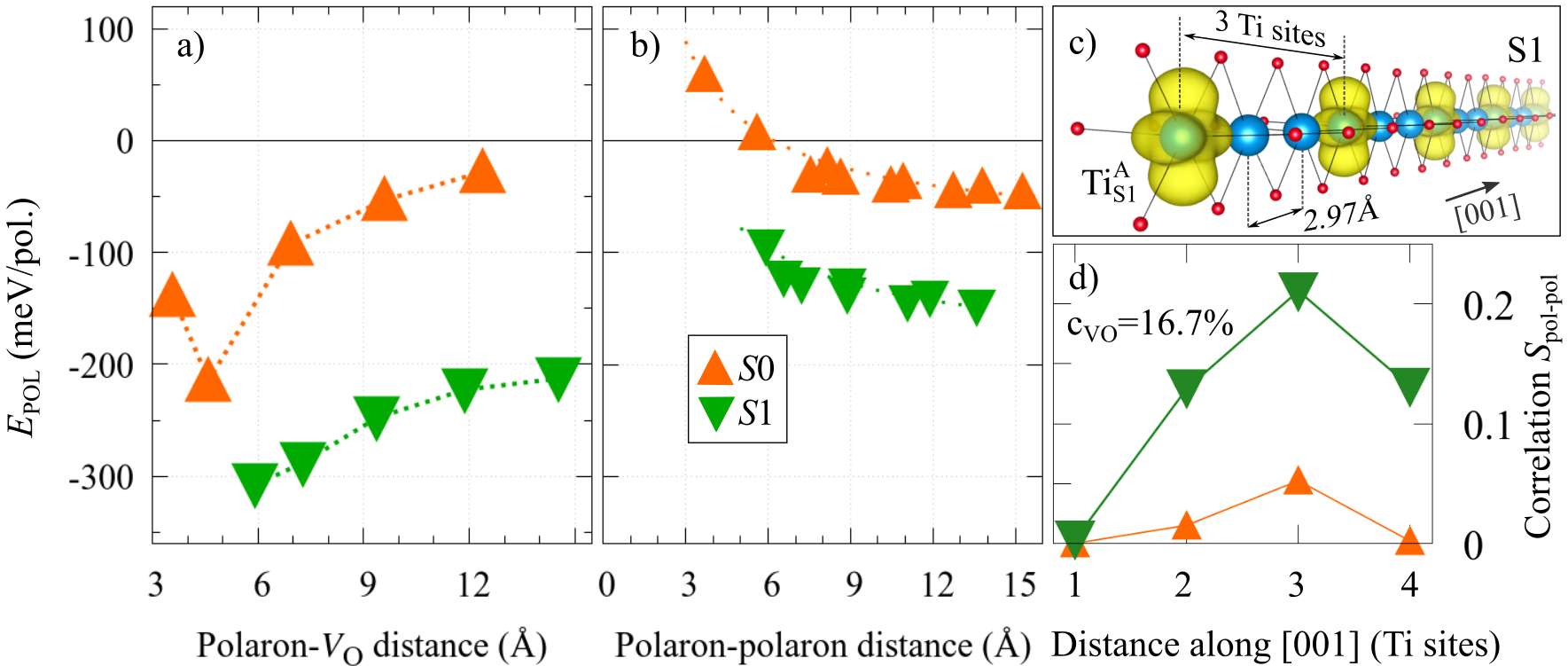}
\caption[Statistics of polarons on rutile \ch{TiO2}(110)]{\textbf{Statistics of polarons on rutile \ch{TiO2}(110):} (a) polaron-\VO\ and (b) polaron-polaron interactions, (c) sketch of the optimal 1$\times$3 polaron pattern in the sub-surface of rutile TiO$_2$ at the critical \VO\ concentration (16.7\%), and (d) the spatial correlation function.
Adapted from Ref.~\cite{Reticcioli2017c}.
}
\label{fig:stat}
\end{figure}

\emph{Statistical analysis: polaron-polaron and polaron-vacancy interaction}. 
Another relevant aspect to understand the physics of polarons is the Coulomb-like polaron-polaron and polaron-defect (in this case \VO) interaction. 
Being positively charged, a \VO\ acts as an attractive center for the small polarons and, from the structural point of view, renders the local structure more flexible and reduces the structural cost \EST\ to distort the lattice.
Conversely, the polaron-polaron interaction is repulsive and particularly effective at small distances.
The relative distance among polarons and between polarons and \VO\ considerably affects the energy and the degree of localization of the characteristic in-gap polaronic states.
For instance, at short distance the interaction between polarons can be strong enough to split the polaron levels and lead to the onset of double in-gap peaks well separated in energy~\cite{Reticcioli2018a}.
As mentioned in the methodology section, an useful tool to decipher these intricate interactions is the statistical analysis of the polaron energies in the explored configuration space.
The results for polarons in rutile TiO$_2$ are summarized in Fig.~\ref{fig:stat}.

Figures~\ref{fig:stat}(a-b) clearly show that the polaron energy decreases with increasing polaron-polaron distance and with decreasing polaron-\VO\ distance.  
Polaron pairs at nearest-neighbor Ti sites along [001] (\ie\ a polaron-polaron distance of about 3~\AA) is very unstable, due to the strong polaron-polaron repulsion. This Coulomb repulsion is enhanced by the overlap of the polaron charges.
In fact, only 70\% of the excess charge is confined at the trapping site, while the remaining charge is spread around the surrounding atoms and hinders nearest-neighbor polaronic configurations.
For the sake of clarity, it should be noted that the polaron-polaron curve shown in Fig.~\ref{fig:stat}(b) is obtained for a perfect crystal without any \VO, in order to decouple the polaron-polaron interaction from the polaron-\VO\ interaction.
In defective samples, the optimal polaronic configuration is the result of the balance between these two opposite effects: within a simplified picture, one can conclude that polarons try to maximize their mutual distance and to minimize the polaron-\VO\ distance.

Clearly the most favorable polaron arrangement and its energetics is concentration dependent. 
In rutile TiO$_2$, the highest \VO\ concentration experimentally achievable is $\approx$ 16.7 \%~\cite{Reticcioli2017c}. At low concentrations, polarons occupy $S1$ sites.
By progressively increasing the reduction level, the excess electrons form increasingly more polarons at the most convenient sites in $S1$, finally forming a characteristic 1$\times$3 pattern at the optimal concentration of 16.7\%~\cite{Reticcioli2017c, Shibuya2017b}. This is well reproduced by theory, as exemplified by the peak at a 3-site distance in the correlation function shown in Fig.~\ref{fig:stat}(d). A pictorial view of the 1$\times$3 pattern is shown in Fig.~\ref{fig:stat}(c).
Larger \VO\ concentrations are energetically problematic, since additional polarons in $S1$ would result in strong polaron-polaron repulsion. To solve this instability, the system undergoes a structural reconstruction, as discussed in the following.

\begin{figure}[!ht]
\centering
        \includegraphics[width=1.0\columnwidth,clip=true]{\dirimg 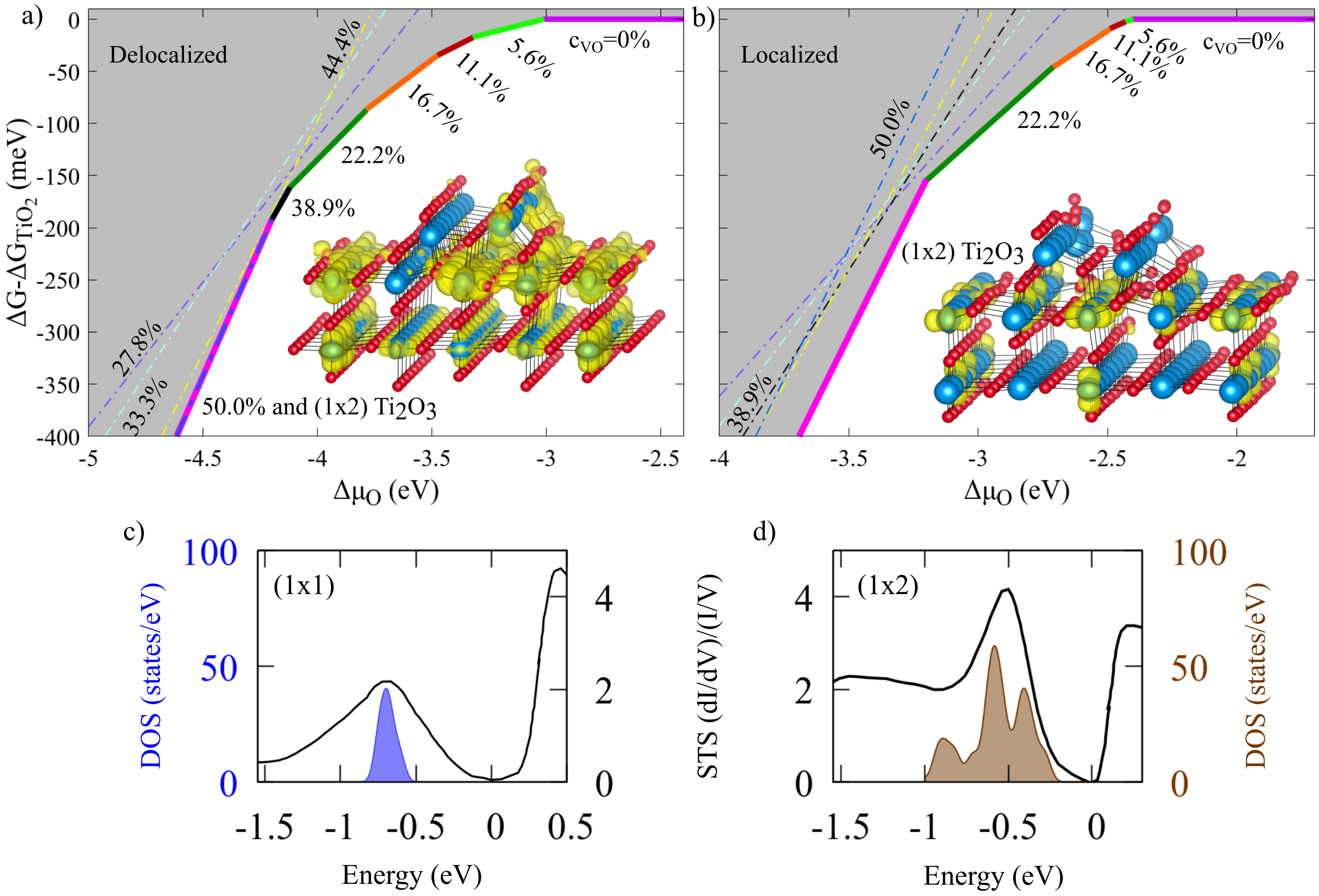}
\caption[Rutile \ch{TiO2}(110) phase diagram]{\textbf{Rutile \ch{TiO2}(110) phase diagram.}
The surface energy of the (1$\times$1) phase with various concentrations of oxygen vacancies (\cVO) and of the (1$\times$2) phase is reported as a function of the oxygen chemical potential $\mu_{\rm O}$, as obtained by delocalized (a) and localized (b) calculations. Panels (c) and (d) display a comparison between the experimental (STS) and simulated (DFT+$U$) polaronic peak in the (1$\times$1) and reconstructed (1$\times$2) surfaces.
This Figure is adapted from Ref.~\cite{Reticcioli2017c}.
}
\label{fig:phasediag}
\end{figure}

\emph{Surface phase diagram (polaron-induced surface reconstruction)}. 
The existence of a critical concentration of polarons determines a limit for the oxygen vacancy formation via O$_{2c}$ removal.
In fact, rutile \ch{TiO2} is able to host up to 16.7\% \VO\ on the surface, which corresponds to a concentration of 33\% polarons in the $S1$ Ti [001] rows (each \VO\ introduces two excess electrons).
However, the system can host larger quantities of excess electrons via a structural reconstruction.
The (1$\times$1) surface undergoes a (1$\times$2) reconstruction, specifically the \ch{Ti2O3} model~\cite{Onishi1994, Reticcioli2017c}, which is able to conveniently handle a larger reduction of the system (50\%), as discussed in Ref.~\cite{Reticcioli2017c}.
This surface reconstruction problem can be described by DFT+$U$ by construction of an appropriate phase stability diagram [see Fig.~\ref{fig:phasediag}(b)], by calculating the surface free energy using standard
\emph{ab initio} atomistic thermodynamics~\cite{Reuter2001}.
The resulting phase diagram is shown in Fig.~\ref{fig:phasediag}.
Close to the critical concentration the free energy of the reconstructed surface is lower than the corresponding unreconstructed one, thereby marking the structural phase transition. Remarkably, by neglecting polaron effects [panel (a), delocalized setup], DFT+$U$ calculations find the (1$\times$2) phase unstable, in disagreement with observations~\cite{Reticcioli2017c}.

It is important to remark that the structural reconstruction is associated with changes of the in-gap states, as shown in Fig.~\ref{fig:phasediag} (c,d): The localized polaron peak, typical of the (1$\times$1) surface (see the corresponding band dispersion in Fig.~\ref{fig:pollevel}), becomes broader in the reconstructed (1$\times$2) phase as a consequence of the higher amount of excess charge localized on the surface.

\begin{figure}[!ht]
\centering
        \includegraphics[width=1.0\columnwidth,clip=true]{\dirimg 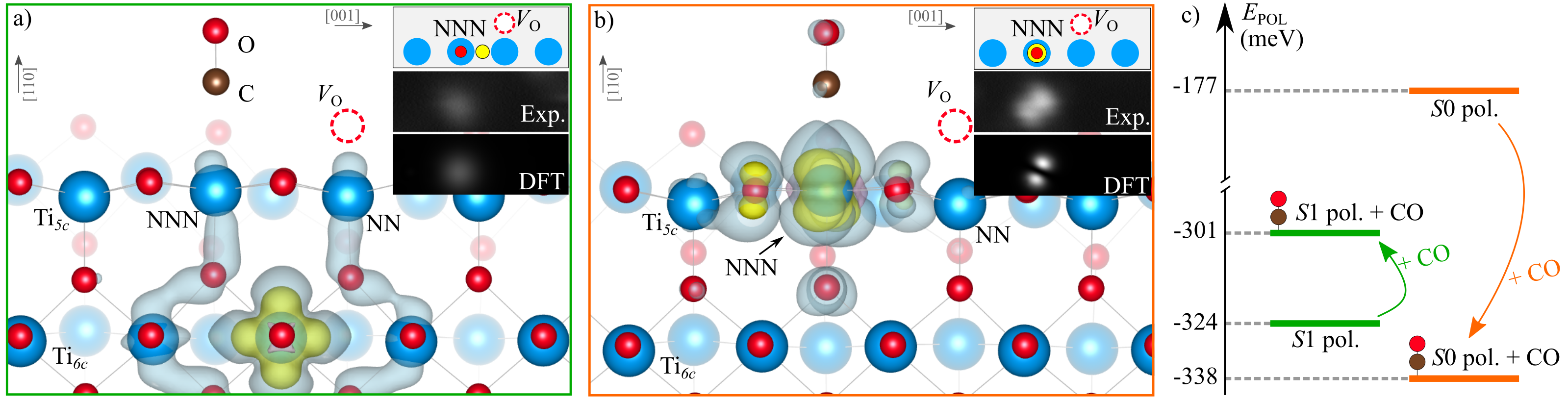}
\caption[CO adsorption on rutile \ch{TiO2}(110)]{\textbf{CO adsorption on rutile \ch{TiO2}(110).}
The CO adsorption at Ti$_{5c}$ sites in presence of sub-surface (a) and surface (b) polarons.
The insets sketch the geometric configuration and report the simulated and experimental STM signatures.
The polaron formation energy perturbed by the CO is compared to the unperturbed case (c).
This figure is adapted from Ref.~\cite{myReticcioliCO}.
}

\label{fig:adsorption}

\end{figure}


\emph{Interaction between polarons and adsorbates}. 
A final paradigmatic case of the importance on polarons on surfaces is the interaction of polarons with adsorbates.
This interaction is essential to decode the initial stage of catalytic reactions at defective surfaces.
It has been shown that even when adsorbates do not transfer any charge carrier to the substrate, polarons 
can affect the energetics and configuration of the adsorption process.
This is the case for CO adsorption on \ch{TiO2}(110)~\cite{myReticcioliCO}, where the polaron stability is altered by the presence of this adsorbate, as shown in Fig.~\ref{fig:adsorption}.

Adsorbed CO shows attractive coupling with polarons on the surface layer, and repulsive interaction with polarons in the subsurface. As a result, upon CO adsorption the most convenient polaron site is not in $S1$ anymore: polarons are attracted onto the surface layer ($S0$) and bind with the CO [Fig.~\ref{fig:adsorption}(c)].
The effect of polarons on $S0$ stands out clearly from the STM analysis of the \ch{TiO2} sample.
A polaron at $S0$, coupled with the CO, is characterized by a bright filled-state STM signal with two lobes tilted with respect [001] [Fig.~\ref{fig:adsorption}(b)], while the $S1$ polaron, right beneath the CO, generates a feeble single-spot, due to a lack of polaronic charge transfer towards the adsorbate [Fig.~\ref{fig:adsorption}(a)].


\section{Summary}

Small polarons can easily form in oxide materials, in particular near the surface where the increased structural flexibility, molecular adsorption and the easy formation of surface defects facilitate the transfer of charge and its coupling with the lattice. Small polarons can be effectively studied using first principles techniques in the framework of DFT.
The main issue to be aware of is that small polarons involve the localization of charge on a particular lattice site. Usually this goes in hand with
a change of the transition metal oxidation state, for instance in TiO$_2$, from Ti$^{4+}$ to Ti$^{3+}$. Such a localization is often
not well described by standard semilocal LDA and GGA functionals because of self-interaction errors.
Therefore it is compulsory to use functionals beyond the semi-local approximation, for instance GGA+$U$ or hybrid functionals.
It goes without saying that such functionals always involve some empiricism, since increasing U or the amount
of exact exchange will favor charge localization and, thus, the formation of polarons. Currently no entirely
satisfactory approach exists giving an unambiguous answer whether polaron formation is favorable or not. But there
are many principles--- such as determination of U using some first principles approaches, or inspection that
the energy versus electron number shows a straight line behavior ---that can guide our choices. Ultimately,
comparison with the experiment, however, remains necessary and mandatory. 

While modeling small polarons, the most important aspects to address are the trapping process, polaron mobility and polaron electronic properties. Using the computational prescriptions presented in this chapter it is possible to explore the configuration space by either guiding the excess charge to localize in selected lattice sites or by performing FPMD calculations. The analysis of the polaron energies (polaron formation energy, structural energy cost and electronic energy gain) allows to identify  the global polaronic minimum and its fundamental energy profile.
In order to fully characterize the polaronic state, it is useful to inspect the degree and extent of the local structural distortions, the mutual interaction among polarons as well as their coupling with defects. This can be done by performing a statistical analysis in terms of the polaron-polaron site correlation function and radial correlation function. A statistical scrutiny of the FPMD data furnishes also important insight on the mobility pattern.
An alternative way to study polaron mobility is Markus theory and the Emin-Holstein-Austin-Mott formalism, which model 
polaron hopping between two sites and permit to discern between adiabatic or non-adiabatic processes.
The electronic characteristics of small polarons are defined by the typical band-gap state formed below the conduction band. For an isolated polaron this state is well localized and typically resides 1~eV below the conduction band. Its specific orbital character (symmetry of the $d$ level where the excess charge is trapped) depends on the local structural and chemical environment.
For interacting polarons the situation is more complicated as spin-exchange effects, Coulomb repulsion and hybridization can significantly modify the energy location and the width of the polaron state.

Apart from their intrinsic significance, polarons can induce dramatic changes in the hosting materials, and can be used as a mean to drive and control different types of transitions, including metal-to-insulator transitions (not discussed here, an example is given in Ref.~\cite{Franchini2009}), structural transitions or adsorption processes. Two examples were discussed in this chapter: (i) polaron-induced surface reconstruction (by increasing the polaron concentration the rutile TiO$_2$(110) undergoes a transition from a (1$\times$1) to a (1$\times$2) surface termination) and (ii) substantial changes of the adsorption energy (the interaction between CO and polarons changes the polaron configuration and affects the CO adsorption scheme). 

Finally, it is important to underline that the quality of the theoretical modeling can be assessed by a direct comparison with experimental data, including EPR (by recognizing the change in the formal valence state of the trapping site; for instance in TiO$_2$ polaron formation changes the Ti valence state from Ti$^{4+}$ to Ti$^{3+}$), spectroscopy (energy level and band dispersion), or STM (by a direct comparison between measured and simulated STM images).

\begin{acknowledgement}
The authors gratefully acknowledge the support of the Austrian Science Fund
(FWF) SFB project VICOM (Grant No. F41) and FWF project POLOX (Grant No. I 2460-N36).
The data and analysis presented in this chapter are the results of years long collaborations on the physics of polarons with the theoretical colleague X. Hao (University of Vienna, now Yanshan University), A. Janotti and C. Van de Walle (University of California, Santa Barbara) and with the experimental collaborators M. Setvin and M. Schmid (Surface Science Group @ TU Wien). Their invaluable help and stimulating ideas were essential to develop the research on polarons discussed in this work. 
\end{acknowledgement}
\bibliographystyle{apsrev4-1-mic1812}
\bibliography{bib-2018-09-27-MOD}

\end{document}